\documentclass[aps,prb,reprint,groupedaddress]{revtex4-2}

\usepackage{graphicx}
\usepackage{amsmath,amssymb}

\begin{document}



\title{Tuning critical field, critical current, and diode effect of narrow thin-film superconductors through engineering inhomogeneous Pearl length}


\author{Takayuki Kubo}
\email[]{kubotaka@post.kek.jp}
\altaffiliation[]{Stay-at-home dad on paternity leave for more than 2.5 years, New York, NY, USA}
\affiliation{High Energy Accelerator Research Organization (KEK), Tsukuba, Ibaraki 305-0801, Japan}
\affiliation{The Graduate University for Advanced Studies (Sokendai), Hayama, Kanagawa 240-0193, Japan}



\begin{abstract}
We explore critical field and critical current behavior in inhomogeneous narrow thin-film superconducting strips. 
Formulations are developed to calculate free energy, critical field, and critical current for strips with inhomogeneous Pearl length distributions. 
Our findings show that inhomogeneities, specifically a shorter Pearl length in the middle of the strip, significantly enhance the critical field $B_{c1}$. 
This has practical implications for achieving complete flux expulsion. 
While narrow strips have traditionally been considered the most effective approach to improve $B_{c1}$ and eliminate trapped vortices, our results suggest that engineered inhomogeneities offer an alternative method to enhance the critical field and improve flux expulsion without reducing strip width, providing greater design flexibility for superconducting devices.
Additionally, we find that for the purpose of increasing the critical current, utilizing an inhomogeneous film with a reduced Pearl length in the middle of the strip is advantageous. 
The enhancement in critical current arises from the current suppression effect at the edges induced by the inhomogeneous distribution of superfluid density. 
Furthermore, we demonstrate that an inhomogeneous film with a left-right asymmetric Pearl length distribution enables control over the nonreciprocity of the critical current, highlighting the potential of engineering inhomogeneous Pearl length distributions to implement devices exhibiting the superconducting diode effect.
Our results provide concrete examples of how manipulating the inhomogeneity of Pearl length can enhance the performance of superconducting devices. 
Various methods such as doping nonuniform impurities or creating a temperature gradient can be employed to implement an inhomogeneous Pearl length distribution.
\end{abstract}

\maketitle


\section{Introduction} \label{introduction}

\begin{figure}[tb]
   \begin{center}
   \includegraphics[width=0.49\linewidth]{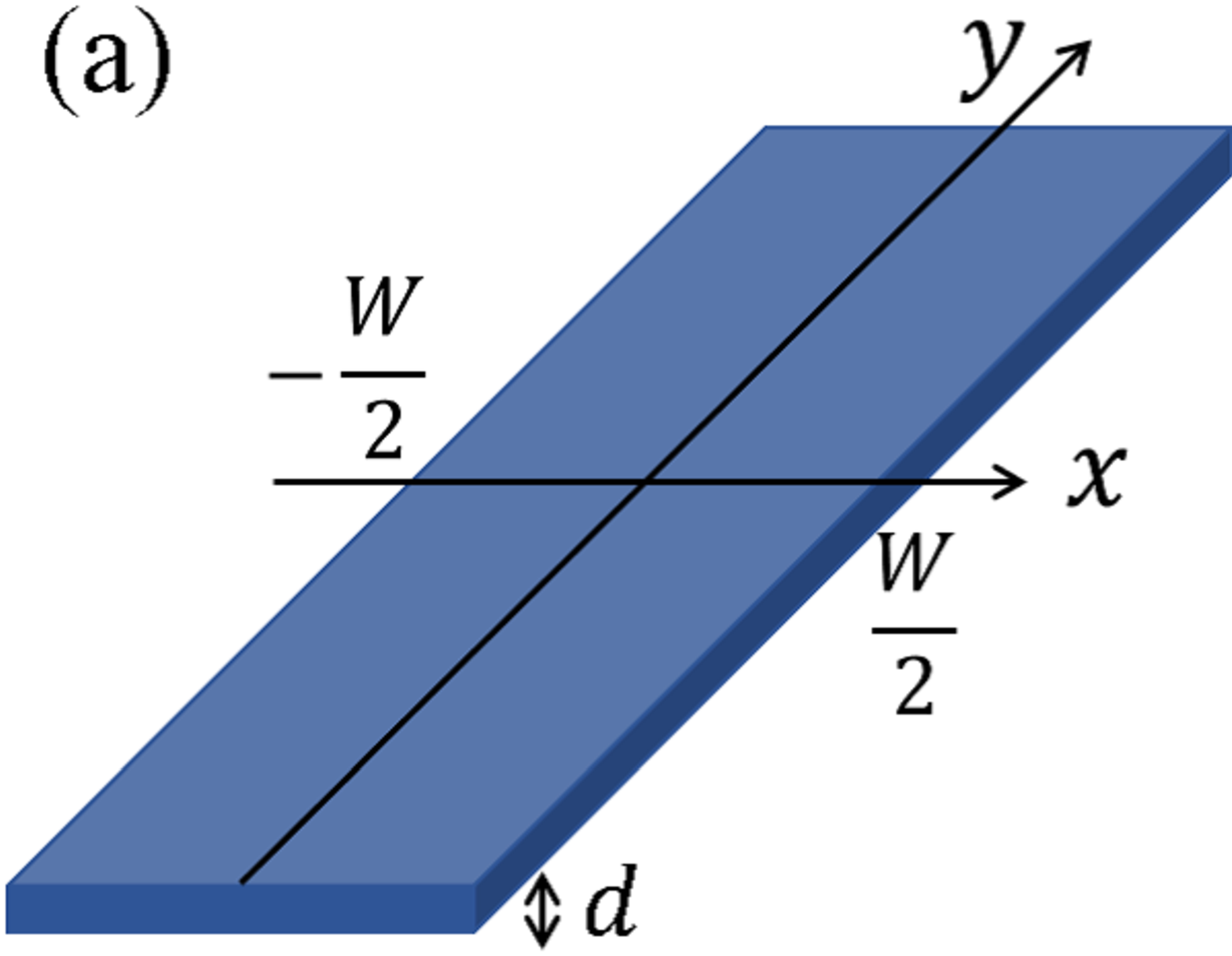}
   \includegraphics[width=0.49\linewidth]{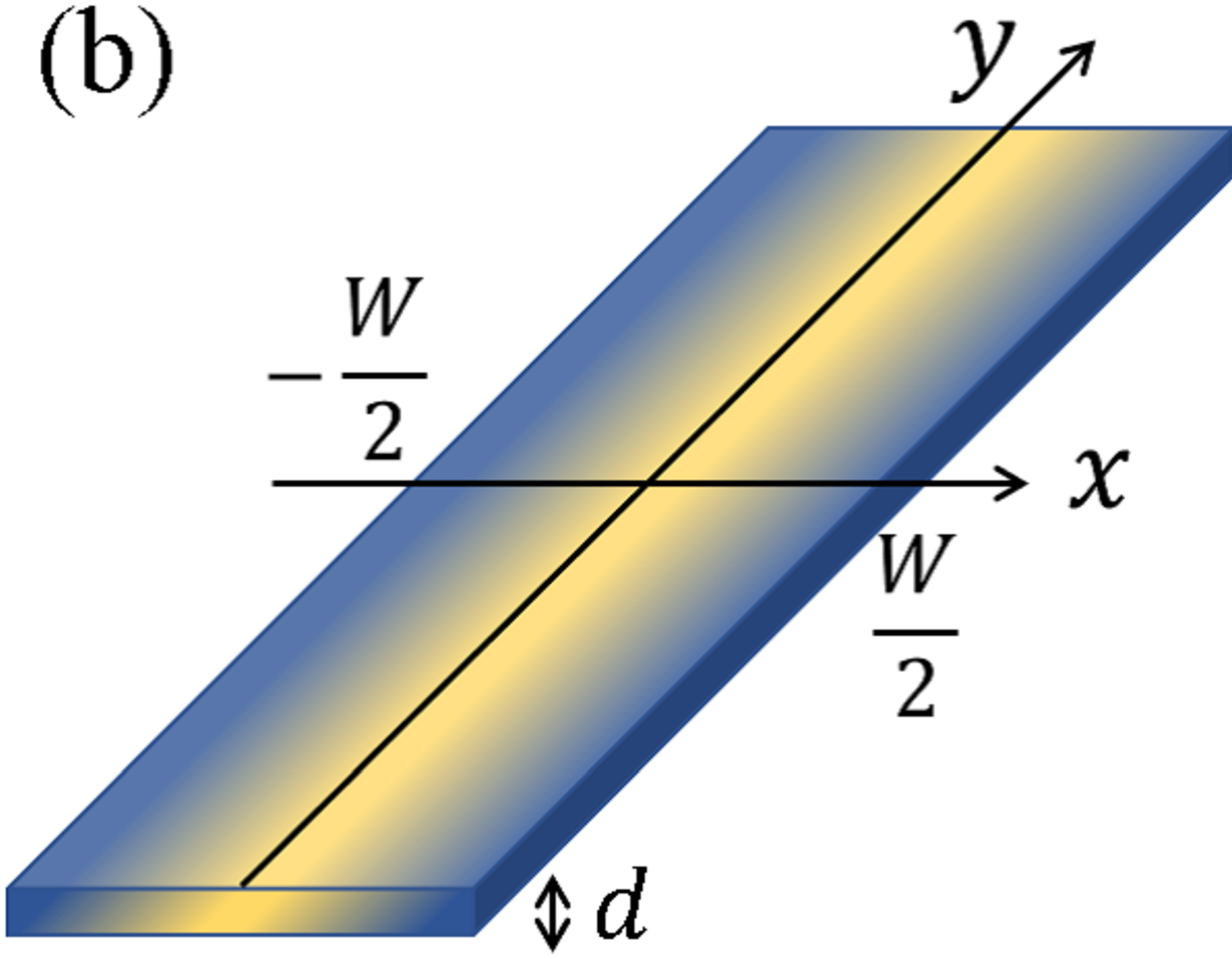}
   \end{center}\vspace{0 cm}
   \caption{
Schematic illustrations of the geometries investigated in the present study. 
The difference in color represents the distribution of Pearl length.  
(a) Homogeneous narrow thin-film strip, and (b) inhomogeneous narrow thin-film strip.
The film thickness $d$ and width $W$ satisfy $d < \lambda$ and $W \ll \Lambda$, respectively.
   }\label{fig1}
\end{figure}

Engineering materials with an inhomogeneous structure, such as optimized impurity profiles and heterostructures, has the potential to enhance the performance of various superconducting devices, including superconducting qubits~\cite{2019_Q_report}, astrophysics detectors~\cite{2017_Klapwijk_Semenov, 2004_Zmuidzinas, 2012_Zmuidzinas}, and particle accelerators~\cite{2017_Padamsee, 2017_Gurevich_SUST}. 
Researchers in the field of superconducting resonators for particle accelerators have actively explored different techniques to optimize impurity-diffusion profiles and improve the overall performance of cavity resonators.
Notably, the development of impurity doping techniques in the 2010s~\cite{2013_Grassellino, 2013_Dhakal, 2017_Grassellino, 2018_Dhakal, 2017_Konomi, 2020_Lechner}, along with the recent introduction of medium temperature baking~\cite{2020_Romanenko, 2020_Posen, 2021_Ito, 2021_He, 2021_Lechner, 2022_Wenskat}, has yielded remarkable advancements, resulting in high-quality factors ranging from approximately $10^{11}$ to $10^{12}$. 
Furthermore, in the 1990s, the oxygen-diffusion technique based on a combination of low-temperature baking and electropolishing~\cite{Saito, Lilje, Ciovati_bake} demonstrated the capability to achieve large microwave amplitudes near the superheating field~\cite{KuboHG, GengHG}, where the screening current density approaches the depairing current density~\cite{1963_Maki, 1980_Kupriyanov, 2012_Lin_Gurevich, 2020_Kubo_1, 2020_Kubo_2, 2022_Kubo}. 
Additionally, thin-film heterostructures formed on bulk niobium have been proposed as a means to enhance the achievable microwave amplitudes further~\cite{2006_Gurevich, 2014_Kubo, 2015_Gurevich, 2017_Liarte_SUST, 2017_Kubo_SUST, 2019_Sauls, 2021_Kubo}. 
Sample tests have demonstrated positive outcomes, validating the effectiveness of the proposed methods~\cite{Tan, 2017_Anne-Marie, 2019_Antoine, ItoSIS, KatayamaSIS, 2021_Lin, snowmass_thin_film}.
Considering these achievements, an intriguing question arises: Can we effectively apply these techniques involving inhomogeneous structures to superconducting thin-film devices? 
To explore this possibility, this article undertakes a theoretical investigation focusing on a narrow thin-film strip with a large in-plane penetration depth characterized by the Pearl length $\Lambda=2\lambda^2/d \gg W$ as a representative system (see also Figure~\ref{fig1}). 
Here, $\lambda$ represents the London depth, $d$ is the thickness of the film (with $d < \lambda$), and $W$ denotes the width of the strip.

One example that highlights this approach is the enhancement of the critical field in narrow thin-film superconducting strips in the perpendicular magnetic field, resulting in the complete expulsion of vortices. Trapped vortices contribute to dissipation in superconducting devices, emphasizing the importance of minimizing vortex trapping for improved resonator performance. 
In homogeneous narrow thin-film strips, vortices are expelled below a nearly material-independent critical field $B_{c1}$ given by $B_{c1} \sim \phi_0/W^2 \ln (W/\xi)$~\cite{Likharev, Martinis, Bronson, 2012_Bulaevskii_Graf_Kogan}, 
where $\phi_0$ is the flux quantum and $\xi$ is the coherence length. 
Experimental studies have provided evidence consistent with these theoretical observations~\cite{Martinis}. 
Consequently, it has been widely believed that the most practical approach to improve $B_{c1}$ and eliminate trapped vortices is through the design of devices with narrow strips.
In this article, we demonstrate that this strong constraint on $B_{c1}$ is limited to homogeneous films and propose an alternative approach to control $B_{c1}$ without modifying their width. 
Our method involves engineering the free energy profile through the introduction of an inhomogeneous distribution of the Pearl length, denoted as $\Lambda(x)$. 
This inhomogeneity can be achieved, for instance, by doping the film with impurities of varying concentrations. 
Our analysis demonstrates that our proposed method is effective in increasing or decreasing $B_{c1}$ as desired.

Furthermore, our method of engineering the free energy profile provides a means to control the critical current, enabling the implementation of a specific class of superconducting diode effect (SDE) characterized by the unequal disappearance of the vortex energy barrier at the edges~\cite{Hou, 2005_Vodolazov, 2022_Suri}. 
This effect is distinct from other SDE mechanisms~\cite{Watanuki, Ando, Wu, Daido, Yuan, Bergeret, Vasenko, Bauriedl}. 
In the presence of a bias current, the Lorentz force acting on a vortex causes the edge barrier to tilt, allowing vortices to penetrate the thin-film strip once the bias current surpasses the critical value $I_c$, where the edge barrier vanishes. 
In homogeneous films, the critical current remains reciprocal when the edge barriers on both sides are equal, regardless of the magnetic field. 
To achieve a nonreciprocal behavior, it is essential to control the film quality, such as the roughness, at both edges of the film~\cite{Hou, 2005_Vodolazov, 2022_Suri} (see, e.g., Refs.~\cite{Aladyshkin, Clem_Berggren, 2015_Kubo} for the impact of roughness on the vortex barrier). 
On the other hand, in the case of inhomogeneous films, the nonuniform distribution of the Pearl length $\Lambda(x)$ not only affects the edge barrier but also influences the distribution of the sheet bias current. 
By engineering an inhomogeneous $\Lambda(x)$, we can design both the edge barrier and the Lorentz force.
Through our research, we demonstrate the feasibility of inducing nonreciprocal behavior in the critical current by implementing a left-right asymmetric distribution of $\Lambda(x)$ and applying an external magnetic field.

The paper is organized as follows. 
In Section~\ref{section_critical_field}, we delve into the investigation of the critical field $B_{c1}$ and its manipulation. 
We start by examining the established findings in a homogeneous narrow thin-film strip, presented in Sec.~\ref{section_Bc1_hmg}. 
This includes the reproduction of well-known results regarding the free energy and $B_{c1}$.
Subsequently, in Sec.~\ref{section_Bc1_inhmg}, we introduce a formulation that allows us to calculate the free energy of an inhomogeneous narrow thin-film strip. 
We then apply this formulation to a specific problem involving an inhomogeneous $\Lambda(x)$, enabling us to determine the free energy and $B_{c1}$ under such conditions.
A noteworthy outcome of this section is the discovery that by engineering an inhomogeneous $\Lambda(x)$, we can effectively increase or decrease the critical field $B_{c1}$. 
In Sec.~\ref{section_critical_current}, our focus shifts towards investigating the influence of an inhomogeneous $\Lambda(x)$ on the critical current. 
Sec.~\ref{section_Ic_hmg} provides a brief overview of the critical current in homogeneous narrow thin-film strips.
In Sec.~\ref{section_Ic_inhmg}, we formulate the critical current ($I_c$) calculation for inhomogeneous strips and study the impact of concrete $\Lambda(x)$ distributions.
Our study uncovers the potential to achieve nonreciprocal critical currents by implementing left-right asymmetric $\Lambda(x)$ distributions in the presence of an external magnetic field. 
This discovery opens new possibilities for designing devices that exhibit the superconducting diode effect.
In Sec.~\ref{discussion}, we discuss the implications of our results.

\section{Critical field} \label{section_critical_field}

\subsection{Free energy and critical field in a {\it homogeneous} narrow thin-film strip} \label{section_Bc1_hmg}

Consider a Pearl vortex in the geometry depicted in Fig.~\ref{fig1} (a), 
which has been extensively studied in the past few decades~\cite{Likharev, Martinis, Bronson, 2012_Bulaevskii_Graf_Kogan, Maksimova, 1994_Kogan, 1998_Clem, 2007_Kogan, Clem_Mawatari, Ilin_Vodo, Kogan_Ichioka}. 
In this subsection, we present an alternative approach to reproduce some well-established results from previous studies, which we will utilize in the subsequent subsections to address more complicated cases: inhomogeneous films.

\subsubsection{Maxwell-London equation for a homogeneous narrow thin-film strip and its solution}

Assuming the presence of a single vortex at ${\bf r}_v=(X,0)$, the Maxwell-London (ML) equation governing the behavior of a homogeneous narrow thin-film strip is given by $B + \mu_0 \lambda^2 ({\rm rot} {\bf j})_z = \phi_0 \delta^{(2)}({\bf r}- {\bf r}_v)$, where $B$ is the magnetic field perpendicular to the film, $\lambda$ is the London depth, ${\bf j}$ is the supercurrent density, and $\phi_0$ is the flux quantum. 
Averaging this equation over the $z$ direction yields $B + (1/2) \mu_0 \Lambda ({\rm rot} {\bf J})_z = \phi_0 \delta^{(2)}({\bf r}- {\bf r}_v)$, where ${\bf J}=\int dz {\bf j}$ is the sheet current, and $\Lambda=2\lambda^2/d$ is the Pearl length. Since we consider a narrow film with $W\ll \Lambda$, the second term dominates the left-hand side. Thus, we obtain the following equation
\begin{eqnarray}
\frac{\mu_0 \Lambda}{2} ({\rm rot} {\bf J})_z = \phi_0 \delta^{(2)}({\bf r}- {\bf r}_v). \label{ML1}
\end{eqnarray}
The boundary condition is
\begin{eqnarray}
J_x (\pm W/2, y) = 0. \label{BC1}
\end{eqnarray}
We adopt the approach of Kogan~\cite{2012_Bulaevskii_Graf_Kogan, 1994_Kogan, 2007_Kogan, Kogan_Ichioka} and reformulate Eqs.~(\ref{ML1}) and (\ref{BC1}).
Because ${\rm div}{\bf J}=0$, we can introduce the stream function $\Psi_{X}(x,y)$, or alternatively, its dimensionless version $\psi_{X}(x,y)=(\mu_0 \Lambda/2\phi_0) \Psi_X(x,y)$, by
\begin{eqnarray}
{\bf J} = {\rm rot} (\Psi_X {\bf \hat{z}}) =  \frac{2\phi_0}{\mu_0 \Lambda} {\rm rot} (\psi_X {\bf \hat{z}}) 
=\frac{2\phi_0}{\mu_0 \Lambda}
\begin{pmatrix}
\partial_y \psi_X\\
-\partial_x \psi_X\\
0
\end{pmatrix} 
. \label{StreamFunction} 
\end{eqnarray}
Here, the subscript $X$ indicates the position of the vortex at ${\bf r}_v=(X,0)$.
Then, Eqs.~(\ref{ML1}) and (\ref{BC1}) reduce to
\begin{eqnarray}
&& -\nabla^2 \psi_X(x,y) = \delta^{(2)}({\bf r}- {\bf r}_v) , \label{ML2} \\
&& \psi_X(\pm W/2, y) = 0.  \label{BC2}
\end{eqnarray}
The previous studies~\cite{1994_Kogan, 1998_Clem, 2007_Kogan, Kogan_Ichioka} have addressed the solutions of Eqs.~(\ref{ML2}) and (\ref{BC2}) through the use of the method of images and conformal mapping.
In this section, we present an alternative approach for solving Eqs.~(\ref{ML2}) and (\ref{BC2}) using a more direct method.
By employing a Fourier transform of the $y$ coordinate into the $k$ space, we obtain:
\begin{eqnarray}
&& \biggl( -\frac{\partial^2}{\partial x^2} + k^2 \biggr) \tilde{\psi}_X(x,k) = \delta(x-X) , \label{ML3} \\
&& \tilde{\psi}_X (\pm W/2, k) =0 .\label{BC3}
\end{eqnarray}
Here, 
\begin{eqnarray}
\psi_X(x,y) = \int_{-\infty}^{\infty} \frac{dk}{2\pi} \tilde{\psi}_X(x,k) e^{i k y} . \label{Fourier_s}
\end{eqnarray}
Then, Eqs.~(\ref{ML3}) and (\ref{BC3}) are the equation of Green's function for Helmholtz equation in one dimensional box, whose solution is given by
\begin{eqnarray}
\tilde{\psi}_X (x, k) = \frac{\cosh k(W - |x-X|) -\cosh k(x+X)}{2k \sinh kW} . \label{tilde_s}
\end{eqnarray}
Thus, the dimensionless stream function $\psi_X(x,y)$ can be obtained from Eqs.~(\ref{Fourier_s}) and (\ref{tilde_s}).
Notably, the integrand in these equations have poles at $kW = \pm i\pi m$ ($m=1, , 2, , 3,, \dots$).
To evaluate the integrals, we sum the residues at the poles in the upper and lower planes for $y\ge 0$ and $y\le 0$, respectively.
This procedure yields the following expression:
\begin{eqnarray}
\psi_X(x,y) &=& \frac{1}{2\pi} \sum_{m=1}^{\infty} \frac{(-1)^m}{m} \biggl[ 
\cos \biggl\{ m\pi \biggl( 1- \frac{|x-X|}{W} \biggr)\biggr\} \nonumber \\
&&-\cos \biggl( m \pi \frac{x+X}{W} \biggr) \biggr] e^{-m\pi |y|/W} . \label{s}
\end{eqnarray}
Although Eq.~(\ref{s}) may appear complex at first glance, it readily reproduces several well-known formulas, as shown below.

\subsubsection{Self-energy, magnetic moment, and free energy}

\begin{figure}[tb]
   \begin{center}
   \includegraphics[height=0.49\linewidth]{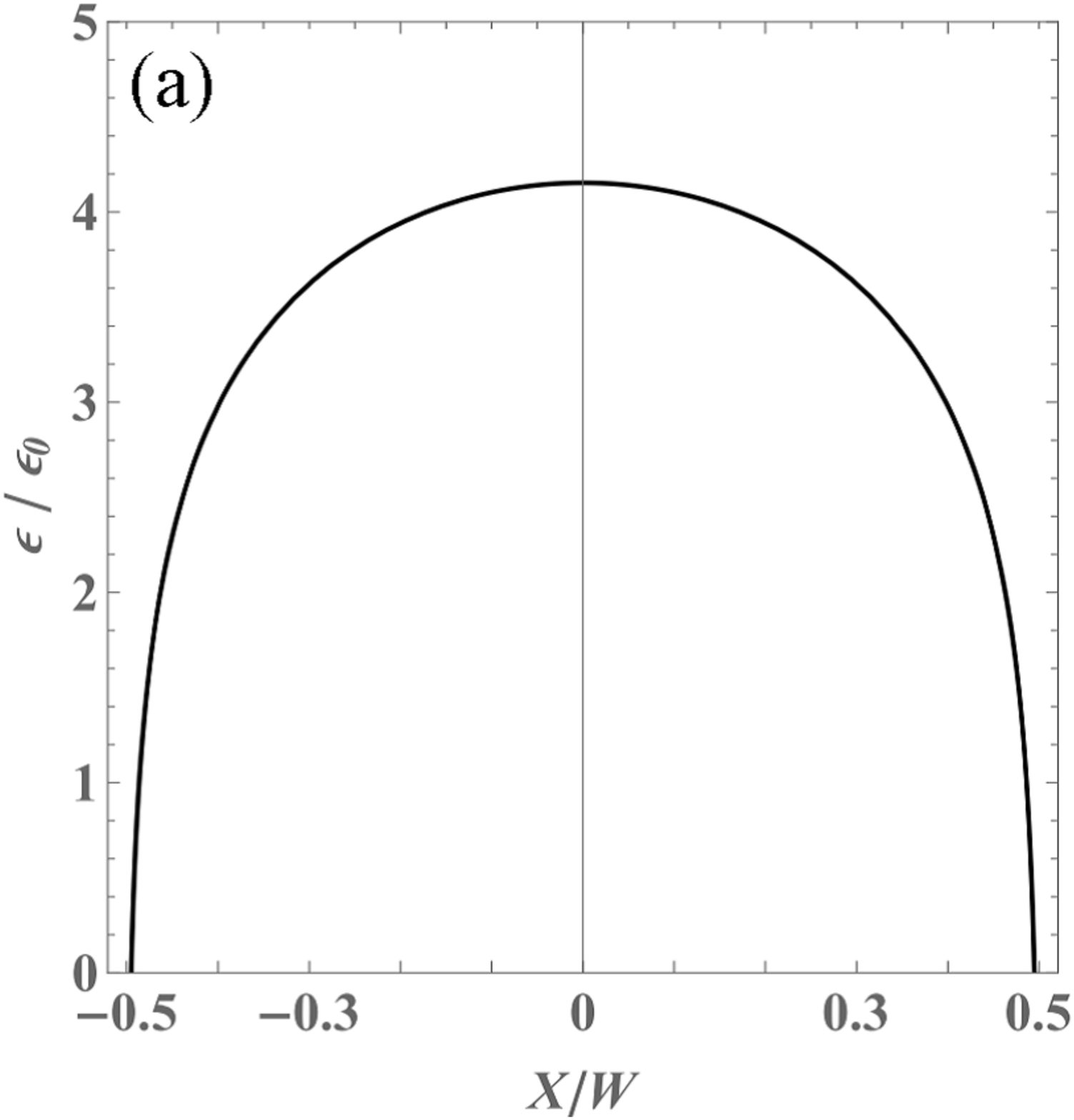}
   \includegraphics[height=0.49\linewidth]{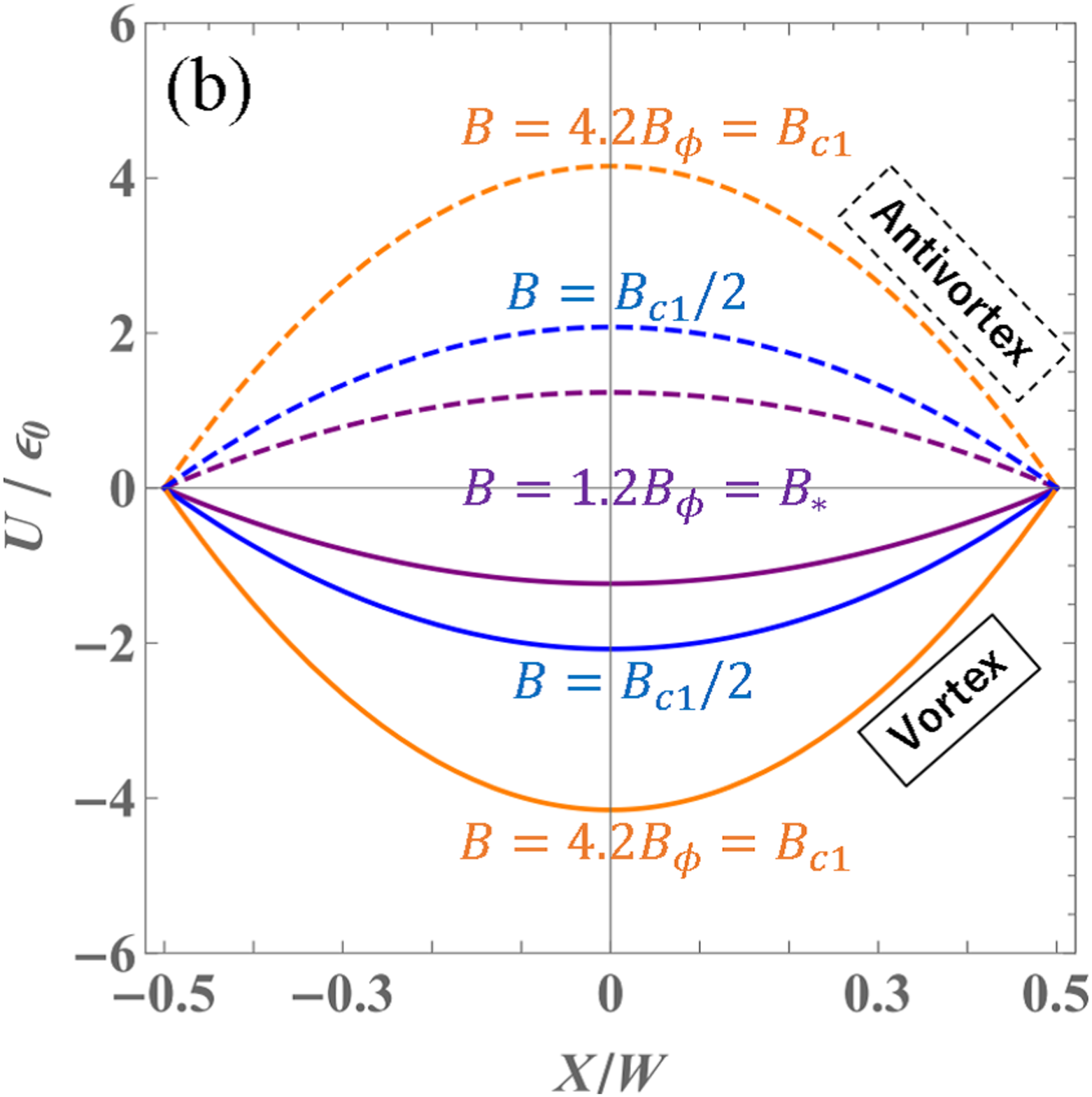}
   \includegraphics[height=0.49\linewidth]{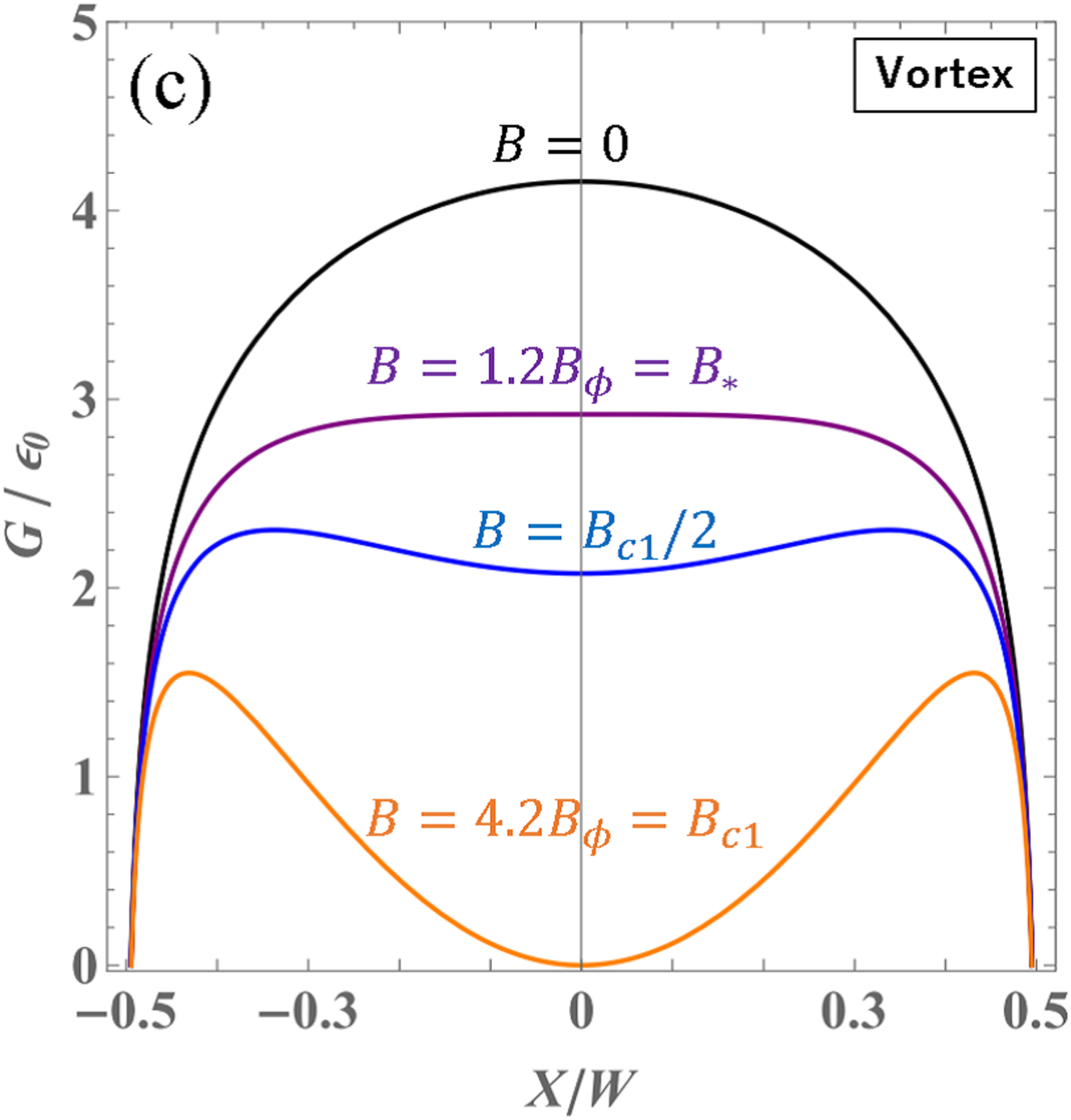}
   \includegraphics[height=0.49\linewidth]{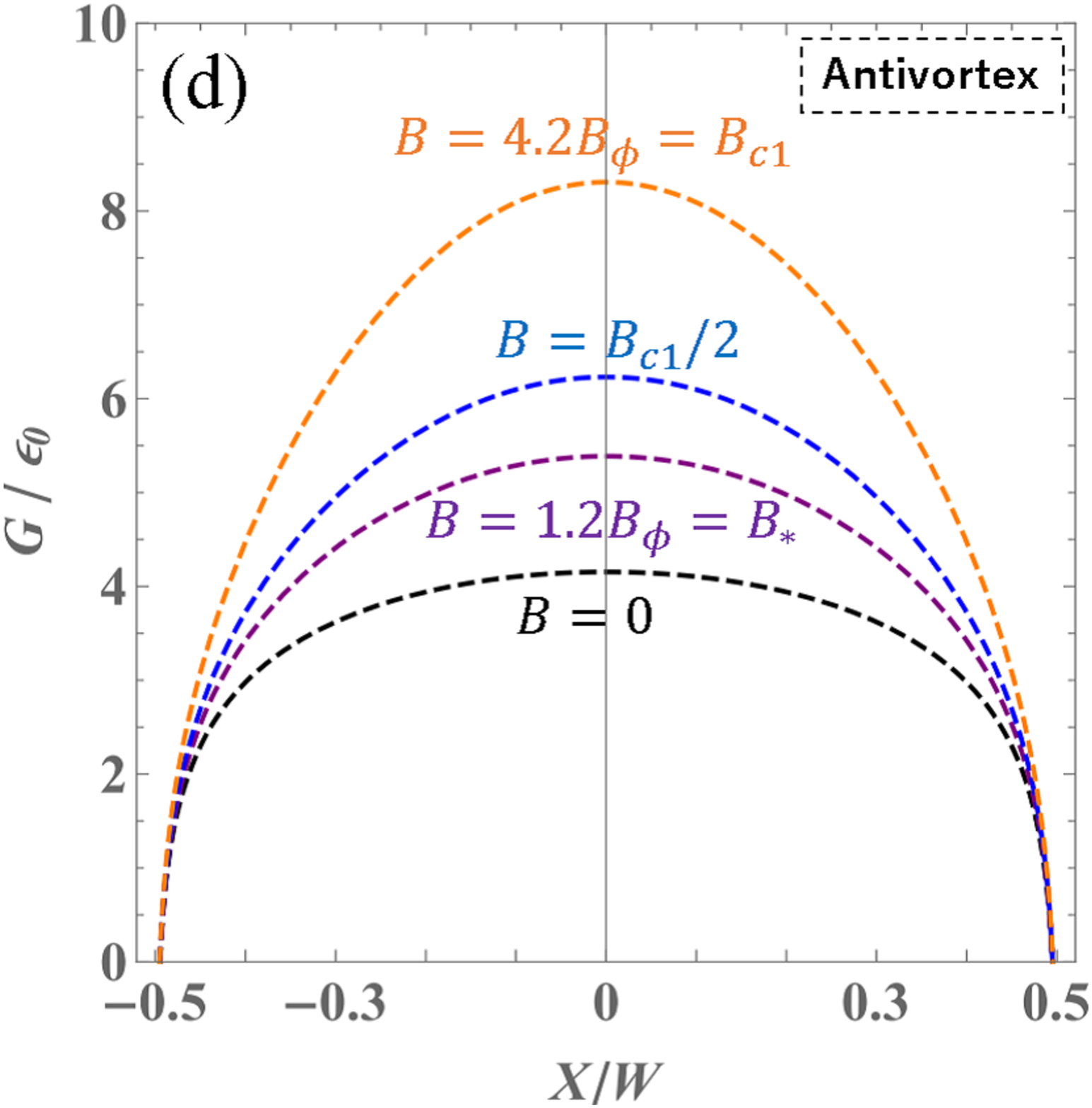}
   \end{center}\vspace{0 cm}
   \caption{
(a) Self energy $\epsilon$ of a vortex or an antivortex as a function of the vortex position ${\bf r}_v=(X, 0)$.
(b) Magnetic potential $U(X)=-\mu_z B$ for a film with a vortex (solid curve) and an antivortex (dashed curve).
(c) Total energy $G(X)$ of a film with a vortex at $x=X$.
(d) Total energy $G(X)$ of a film with an antivortex at $x=X$.
All calculations in the figures are performed with $\xi/W=0.01$.
   }\label{fig2}
\end{figure}

As derived by Kogan~\cite{1994_Kogan, 2007_Kogan}, 
the self-energy of a Pearl vortex in a homogeneous thin-film is given by 
\begin{eqnarray}
\epsilon 
= \frac{\phi_0}{2} \Psi_X(X_{\xi}, 0) 
= \frac{\phi_0^2 \psi_X(X_{\xi}, 0)}{\mu_0\Lambda} , 
\end{eqnarray}
Here, the standard cutoff is applied: ${\bf r}_v=(X,0) \to (X_{\xi}, 0)$ and $X_{\xi}=X+\xi$. 
In Fourier space, we can express this as
\begin{eqnarray}
\epsilon 
= \frac{\phi_0^2}{\mu_0\Lambda}\int_{-\infty}^{\infty}\frac{dk}{2\pi} \tilde{\psi}_X(X_{\xi}, k) . \label{self_energy_formula}
\end{eqnarray}
Using the formula $\sum_{m=1}^{\infty} (-1)^m \cos(2m \alpha)/m = -\ln 2- \ln (\cos \alpha)$, 
we find~\cite{2012_Bulaevskii_Graf_Kogan, 1994_Kogan, 1998_Clem, Bronson, 2007_Kogan, Tafuri, Clem_Mawatari}
\begin{eqnarray}
\epsilon &=& \frac{\phi_0^2}{2\pi \mu_0 \Lambda} \ln \frac{\cos (\pi X / W)}{\cos(\pi/2-\pi\xi/2W)} \\
&\simeq &  \frac{\phi_0^2}{2\pi \mu_0 \Lambda} \ln \biggl( \frac{2W}{\pi\xi} \cos \frac{\pi X}{W} \biggr) . \label{self_energy}
\end{eqnarray}
The assumption $\xi/W \ll 1$ is utilized in the last line of the equation.
Eq.~(\ref{self_energy}) represents the well-known formula for the self-energy of a Pearl vortex in a homogeneous narrow thin-film strip.
It is worth noting that, in other papers, the strip is often defined over the interval $[0,W]$ rather than our interval of $[-W/2, W/2]$.
To obtain the same expression, we must transform $X \to X + W/2$.
Figure~\ref{fig2}(a) shows the function $\epsilon$ of a vortex (antivortex) positioned at $x=X$.

When a vortex exists in the film, it generates a magnetic moment $\mu_z$ within the film, 
which contributes to the magnetic potential $U=-\mu_z B$. 
The magnetic moment can be calculated from
\begin{eqnarray}
\mu_z(X) = \frac{1}{2} \int \!\!d^2 r ( {\bf r} \times {\bf J} )_z
= \frac{2\phi_0}{\mu_0\Lambda} \int_{-W/2}^{W/2}\!\!\!\! dx \tilde{\psi}_X(x,0) . \label{muz_formula}
\end{eqnarray}
Performing a similar calculation for a film containing an antivortex is straightforward. 
Then, we obtain~\cite{2012_Bulaevskii_Graf_Kogan, 1994_Kogan, 1998_Clem, Bronson, Clem_Mawatari}
\begin{eqnarray}
\mu_z(X) = \pm \frac{\phi_0 W^2}{4\mu_0 \Lambda} \biggl( 1- \frac{4X^2}{W^2} \biggr) , \label{muz}
\end{eqnarray}
for a film with a vortex and an antivortex, respectively. 
It should be noted that,  by replacing $X \to X + W/2$, 
Eq.~(\ref{muz}) can be transformed into the same expression as the papers that consider the strip defined over the interval $[0,W]$. 
Fig.~\ref{fig2} (b) illustrates the behavior of $U=-\mu_z B$ when a vortex (antivortex) is situated at ${\bf r}_v=(X, 0)$.
As $B$ increases (where $B>0$), $U=-\mu_z B$ is stabilized for a vortex and destabilized for an antivortex.

The free energy of the system in the external magnetic field ${\bf B}=(0,0,B)$ is given by~\cite{2012_Bulaevskii_Graf_Kogan, 1994_Kogan, 1998_Clem, Bronson, Clem_Mawatari}
\begin{eqnarray}
&&G(X) = \epsilon(X) - \mu_z(X) B \nonumber \\
&&= \frac{\phi_0^2}{2\pi \mu_0 \Lambda} \ln \biggl( \frac{2W}{\pi\xi} \cos \frac{\pi X}{W} \biggr) 
\mp \frac{\phi_0 W^2 B}{4\mu_0 \Lambda}\biggl( 1- \frac{4X^2}{W^2} \biggr) .
\end{eqnarray}
for a film with a vortex and an antivortex, respectively. 
Introducing 
\begin{eqnarray}
&&\epsilon_0=\frac{\phi_0^2}{2\pi \mu_0 \Lambda}, \\
&&B_{\phi} = \frac{2\phi_0}{\pi W^2} , 
\end{eqnarray}
we have
\begin{eqnarray}
\frac{G(X)}{\epsilon_0} 
= \ln \biggl( \frac{2W}{\pi\xi} \cos \frac{\pi X}{W} \biggr) 
\mp \frac{B}{B_{\phi}} \biggl( 1-4\frac{X^2}{W^2} \biggr) .
\end{eqnarray}
Shown in Fig.~\ref{fig2} (c) is $G$ for a thin-film strip containing a single vortex. 
The vortex state is initially unstable at $B=0$, but becomes stable as $B$ increases. 
As $B$ exceeds a critical value $B_*$ obtained from the condition $d^2G/dX^2=0$ at $X=0$, 
the center of the strip becomes a metastable minimum. 
Here, $B_*$ is given by $B_* = (\pi^2/8)B_{\phi} = \pi \phi_0/4W^2$~\cite{Martinis, Maksimova}, 
where $\phi_0$ is the flux quantum. 
Further increasing $B$ stabilizes the metastable minimum. 
At the critical field $B_{c1}$, which is obtained from the condition $G(0)=0$, 
the center of the film becomes the global minimum. 
$B_{c1}$ is given by~\cite{Likharev, Martinis, Bronson, 2012_Bulaevskii_Graf_Kogan}
\begin{eqnarray}
B_{c1} = B_{\phi}\ln \frac{2W}{\pi\xi}= \frac{2\phi_0}{\pi W^2} \ln \frac{2W}{\pi\xi} . \label{Bc1_hmg}
\end{eqnarray}
The edge barrier disappears at the vortex entrance field $B_v$, 
which is obtained from the condition $dG/dX=0$ at th edges. 
$B_v$ is given by $B_v=(W/4\xi)B_{\phi}=\phi_0/2\pi\xi W$~\cite{Maksimova}, but is not shown in Figure~\ref{fig2}.
Note here that $B_v \gg \{ B_{c1}, B_* \}$ for $\xi/W \ll 1$ (e.g., $B_*=1.2B_{\phi}$, $B_{c1}=4.8B_{\phi}$, and $B_v=50B_{\phi}$ for $\xi/W=0.005$).
On the other hand, $G$ of a film including an antivortex is not stabilized by increasing $B (>0)$ as shown in Fig.~\ref{fig2} (d) but stabilized by $B<0$.

One of the remarkable properties of a vortex is that the critical field $B_{c1}$ is largely independent of material parameters. 
For example, reducing the coherence length $\xi$ from $\xi/W=0.01$ to $\xi/W=0.005$ by changing the circuit material only increases $B_{c1}$ by a factor of 1.17. 
Therefore, if one desires to increase $B_{c1}$, a more realistic option is to decrease the strip width~\cite{Martinis} as long as we use a homogeneous film.
In the following, it will be demonstrated that this constraint can be avoided by utilizing a inhomogeneous film.

\subsection{Free energy and critical field in an {\it inhomogeneous} narrow thin-film strip} \label{section_Bc1_inhmg}

\subsubsection{Formulation} \label{section_formulation}

Consider the geometry depicted in Fig.~\ref{fig1} (b), where $\Lambda$ is not uniform and depends on $x$. 
In this case, Eq.~(\ref{ML1}) is not applicable, and we must use the following equation (see, e.g., Ref.~\cite{Evetts})  
\begin{eqnarray}
\frac{\mu_0}{2} \Bigl[{\rm rot} \{ \Lambda(x) {\bf J}({\bf r}) \} \Bigr]_z = \phi_0 \delta^{(2)}({\bf r} - {\bf  r_v}) . \label{inhmg_ML}
\end{eqnarray}
Writing $\Lambda(x) = \Lambda_0 F(x)$ and introducing the potential ${\bf J}={\rm rot} (\Psi_X{\bf \hat{z}}) = (2\phi_0/\mu_0\Lambda_0) {\rm rot} (\psi_X{\bf \hat{z}})$, we get 
\begin{eqnarray}
\biggl[ -F(x) \nabla^2  - F'(x) \frac{\partial}{\partial x} \biggr] \psi_X(x,y)  = \delta^{(2)}({\bf r} - {\bf  r_v}) . \label{inhmg_ML3}
\end{eqnarray}
It is worth noting that when $F=1$, we have a uniform $\Lambda$, 
which reproduces Eq.~(\ref{ML2}). 
The Fourier transform of Eq.~(\ref{inhmg_ML3}) is 
\begin{eqnarray}
\biggl( -F \frac{\partial^2}{\partial x^2} + k^2 F  -F' \frac{\partial}{\partial x}\biggr) \tilde{\psi}_X(x,k)  = \delta (x - X) . \label{inhmg_ML4}
\end{eqnarray}
We obtain the solution of Eqs.~(\ref{inhmg_ML3}) or (\ref{inhmg_ML4}) below.

We begin by considering a special case where $F(x)$ is slowly varying. 
When the typical length scale of the spatial variation of $F$ is much larger than that of $\psi$, 
we can approximate $F$ as quasi-homogeneous. 
Under this assumption, Eq.~(\ref{inhmg_ML4}) reduces to the quasi-homogeneous equation, 
$F(x) (-\partial_x^2 + k^2) \psi_X^{\rm qh} = \delta (x-X)$. 
The solution is $\tilde{\psi}_X^{\rm qh} (x,k) = \tilde{\psi}_X^{\rm hmg}(x,k)/F(X)$. 
Here, $\tilde{\psi}_X^{\rm hmg}$ is the solution for the homogeneous film ($F=1$) and is given by Eq.~(\ref{tilde_s}). 
This assumption holds true in the proximity of the vortex core, hence allowing us to derive
\begin{eqnarray}
&&\tilde{\psi}_X(X_{\xi},k) \to \tilde{\psi}_X^{\rm qh}(X_{\xi},k)  \hspace{0.5cm} ({\rm for}~~kW \gg 1) \label{psi_asymptotic}, \\ 
&& \tilde{\psi}_X^{\rm qh}(X_{\xi},k) = 
\frac{\cosh k(W - \xi) -\cosh 2kX}{2k  F(X) \sinh kW} .
\end{eqnarray}
This observation will be useful in the computation of the vortex self-energy in the following.

The self energy $\epsilon$ can be calculated from the formula given by Eq.~(\ref{self_energy_formula}). 
Although this formula was derived for a superconducting thin-film with uniform $\lambda$, 
it can also be applied to the case where $\lambda$ is nonuniform, as shown in the Appendix~\ref{appendix1}. 
Our task is to integrate $\tilde{\psi}_X(X_{\xi}, k)$ over the $k$ space, and to facilitate convergence of the integral, we employ a trick. 
Specifically, we split the integrand into two parts: $\tilde{\psi}_X(X_{\xi}, k)=[\tilde{\psi}_X(X_{\xi}, k) - \tilde{\psi}_X^{\rm qh}(X_{\xi}, k)] + \tilde{\psi}_X^{\rm qh}(X_{\xi}, k)$. 
This yields
\begin{eqnarray}
&&\frac{\epsilon(X)}{\epsilon_0} 
= L_0(X) + L_1(X) , \label{self_energy_inhmg} \\
&&L_0 = \int_{-\infty}^{\infty}\!\!\! dk \tilde{\psi}_X^{\rm qh} (X_{\xi},k) 
= \frac{1}{F(X)} \ln \biggl[ \frac{2W}{\pi \xi}\cos\frac{\pi X}{W} \biggr], \\
&&L_1 = \int_{-\infty}^{\infty}\!\!\! dk \biggl[ \tilde{\psi}_X(X_{\xi},k) - \tilde{\psi}_X^{\rm qh} (X_{\xi},k)  \biggr]  .
\end{eqnarray}
In Section~\ref{section_Bc1_hmg}, we obtained an analytical expression for the integral of $L_0$ by evaluating it explicitly. 
In contrast, the integral of $L_1$ needs to be computed numerically, 
but it converges rapidly thanks to Eq.~(\ref{psi_asymptotic}). 
To perform this numerical evaluation, we introduce a cutoff $k_c$. 
For instance, setting $k_c W \sim 30$-$40$ is sufficient to obtain an accuracy of $\lesssim 1\%$ across a broad range of parameter values used in the subsequent subsections. 
Furthermore, when a vortex approaches the edge, the length scale of the spatial variation of $\tilde{\psi}$ decreases due to the boundary condition, making the quasi-homogeneous contribution $L_0$ dominant. 
This leads to the following equation.
\begin{eqnarray}
\frac{\epsilon(X)}{\epsilon_0} \biggl|_{X\simeq \pm W/2} 
\simeq L_0(X) = \frac{1}{F(X)} \ln \biggl[ \frac{2W}{\pi \xi}\cos\frac{\pi X}{W} \biggr]. 
\label{self_energy_inhmg_edges} 
\end{eqnarray}
The magnetic potential $U=-\mu_z B$ also contributes to the free energy. 
The magnetic moment $\mu_z$ can be obtained from Eq.(\ref{muz_formula}) using $\tilde{\psi}(x,0)$, 
which is the solution of Eq.(\ref{inhmg_ML4}) for $k=0$. 
Therefore, we arrive at
\begin{eqnarray}
&&\mu_z(X) = \frac{2\phi_0}{\mu_0 \Lambda_0} \int_{-W/2}^{W/2}\!\!dx \tilde{\psi}_X (x,0) , \label{MM_inhmg}\\
&&\biggl( -F \frac{\partial^2}{\partial x^2}   -F' \frac{\partial}{\partial x}\biggr) \tilde{\psi}_X (x,0)  = \delta (x - X) . \label{inhmg_ML5}
\end{eqnarray}
The free energy $G=\epsilon-\mu_z B$ reduces to 
\begin{eqnarray}
\frac{G(X)}{\epsilon_0} = \frac{\epsilon(X)}{\epsilon_0}
-\frac{B}{B_{\phi}} \frac{8}{W^2} \int_{-W/2}^{W/2}\!\!\!\! dx \tilde{\psi}_{X}(x,0)
\end{eqnarray}
The critical field $B_{c1}$ can be calculated from the condition $G(0)=0$ or  
\begin{eqnarray}
B_{c1} = \frac{\epsilon(0)}{\mu_z(0)} =\frac{B_{\phi}\int_{-\infty}^{\infty}dk \tilde{\psi}_{X=0}(0,k)}{(8/W^2) \int_{-W/2}^{W/2}dx \tilde{\psi}_{X=0}(x,0) }. \label{bc1_inhmg}
\end{eqnarray}
In order to calculate $G(X)$ and $B_{c1}$, a specific Pearl length distribution is required. 
We address it in the next subsection.

As an example of how to implement an inhomogeneous $\Lambda(x)$ into a film, 
let us consider doping impurities. 
In the dirty limit, we know that $\lambda \propto \sqrt{\Gamma_{\rm imp}}$, 
where $\Gamma_{\rm imp}$ is the impurity scattering rate, 
so we have $\Lambda(x) \propto \Gamma_{\rm imp}(x)$. 
Thus, the impurity profile corresponds to the target $\Lambda(x)$ profile, making this one way to implement it.

Note that impurity doping can affect the coherence length, 
which is given by $\xi(x) \propto 1/\sqrt{\Gamma_{\rm imp}(x)} \propto 1/\sqrt{F(x)}$. 
This variation in $\xi(x)$ appears in the standard cutoff of the logarithmic divergence, $\ln (W/\xi)$, 
which is used in the evaluation of the vortex self-energy in the London theory [see Eqs.~(\ref{self_energy}) and (\ref{self_energy_inhmg})]. 
However, it is well known that this artificial cutoff is a qualitative prescription to evaluate vortex self-energy in the London theory, and we should not take seriously a small variation of $\xi$. 
Moreover, even if we take a variation of $\xi$ into account, 
its impact on $\ln (W/\xi)$ is usually negligible. 
Therefore, in this section, we assume that $\xi$ is constant and independent of $\Lambda(x)$.

\subsubsection{Example: quadratic $\Lambda(x)$} \label{section_quadratic}

\begin{figure}[tb]
   \begin{center}
   \includegraphics[width=0.49\linewidth]{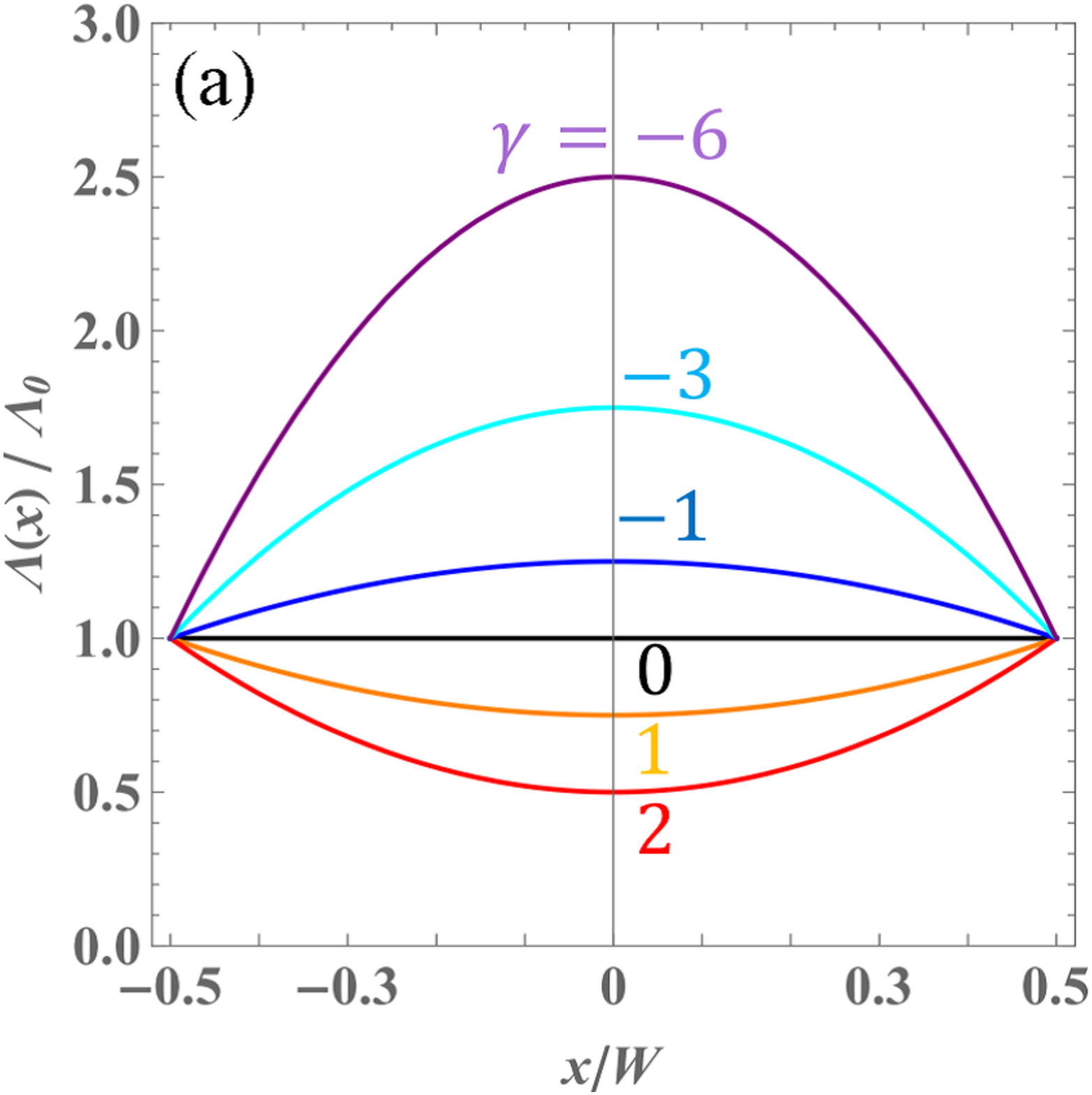}
   \includegraphics[width=0.49\linewidth]{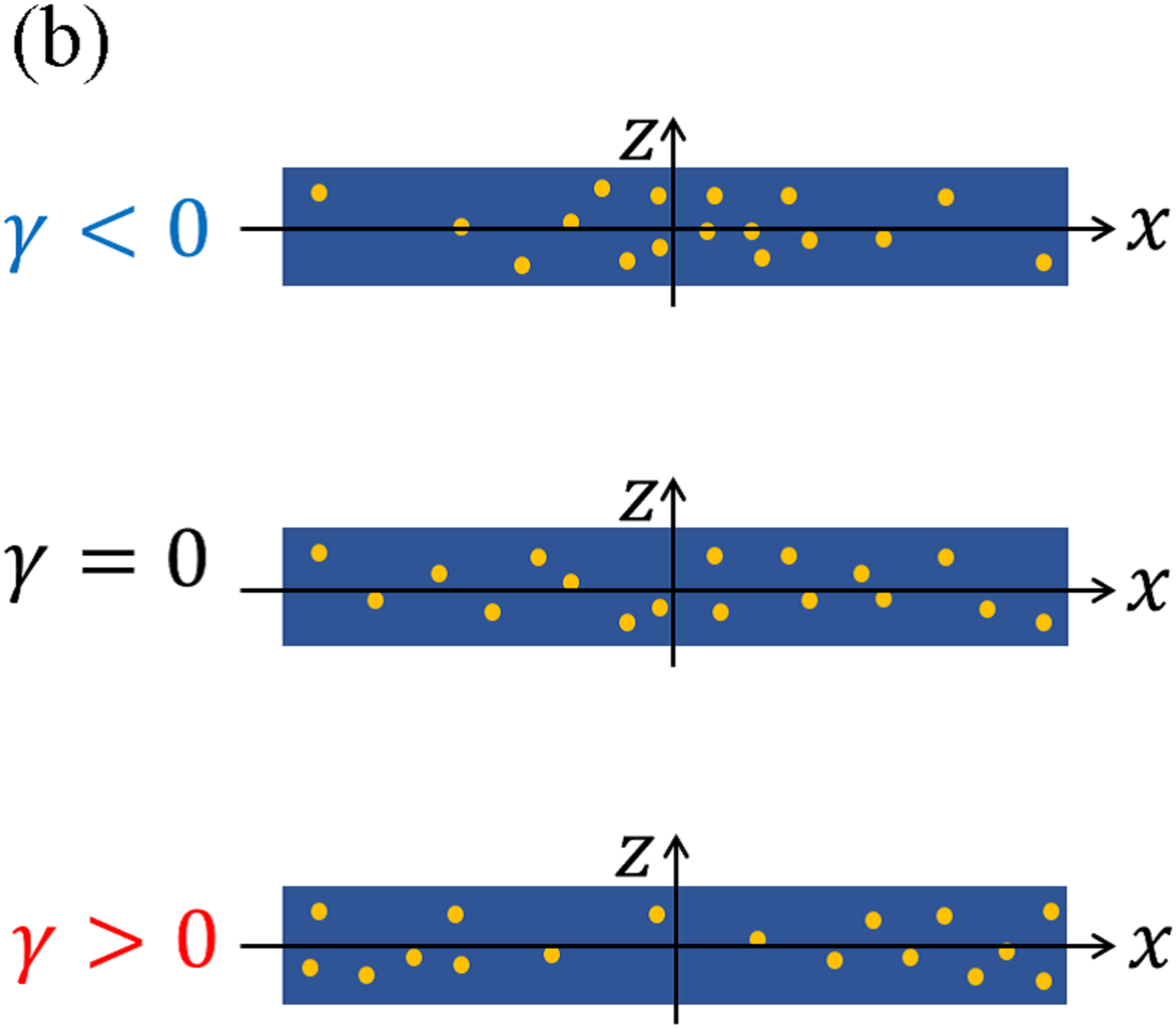}
   \end{center}\vspace{0 cm}
   \caption{
(a) Inhomogeneous distribution of $\Lambda$ given by Eqs.~(\ref{quadratic_Lambda}) and (\ref{quadratic_F}).
(b) Inhomogeneous impurity distributions as examples of methods to implement inhomogeneous $\Lambda$ distributions. 
   }\label{fig3}
\end{figure}

\begin{figure}[tb]
   \begin{center}
   \includegraphics[width=0.49\linewidth]{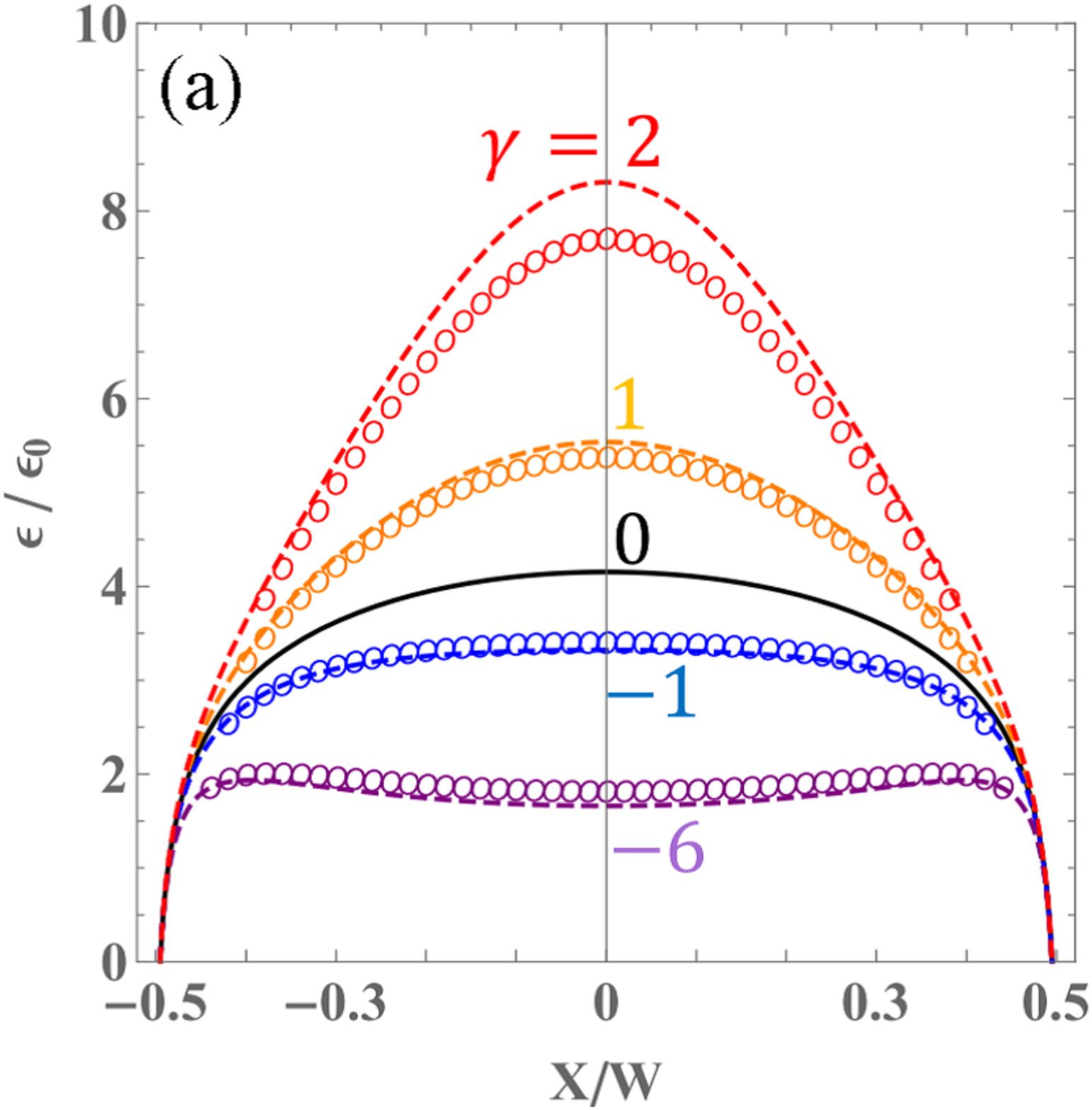}
   \includegraphics[width=0.49\linewidth]{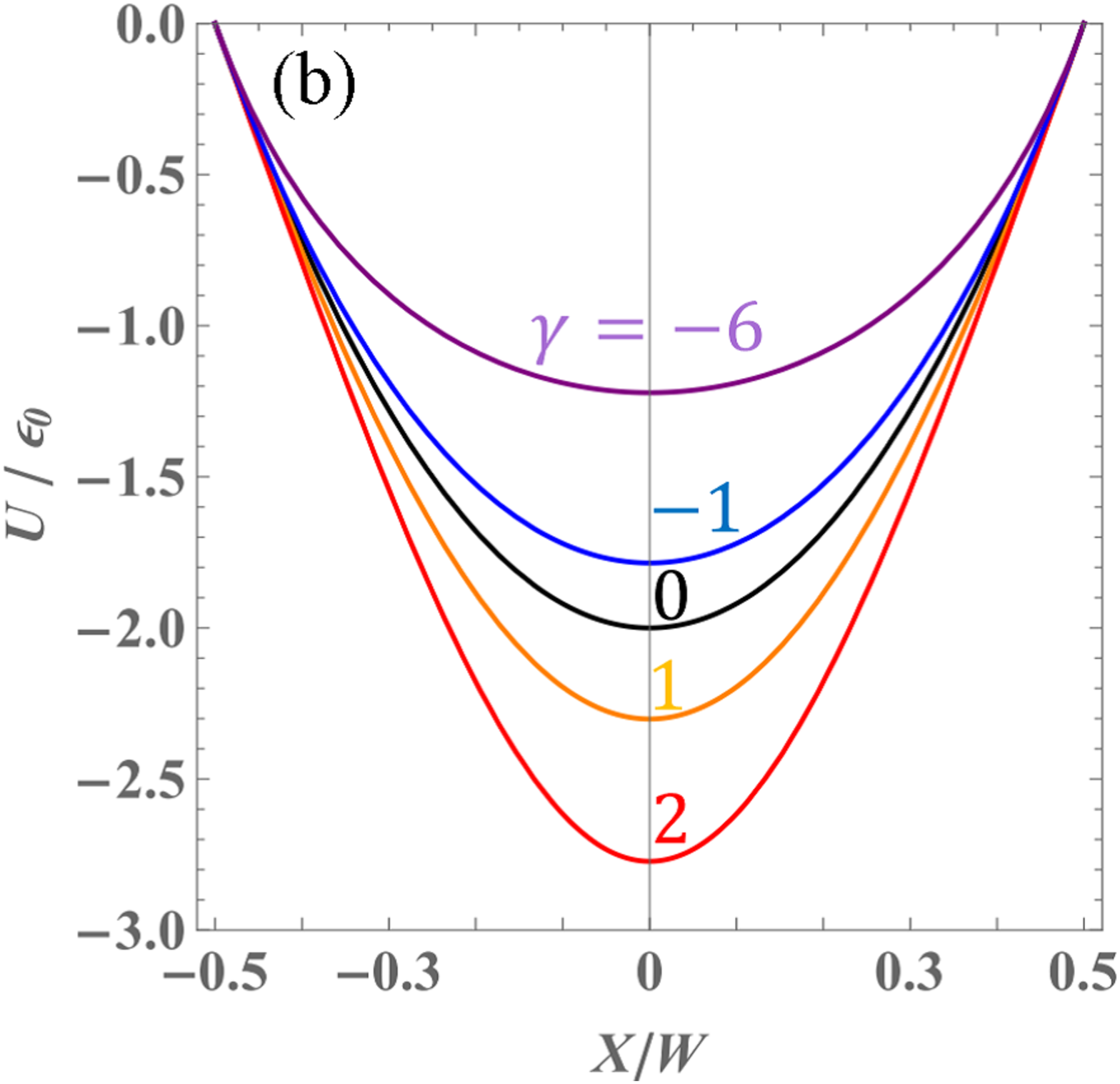}
   \end{center}\vspace{0 cm}
   \caption{
(a) Influence of inhomogeneous $\Lambda$ on the vortex self-energy $\epsilon$. 
The circles represent numerically calculated $\epsilon$ for an inhomogeneous film using Eq.~(\ref{self_energy_inhmg}), 
while the dashed curves are calculated based on Eq.~(\ref{self_energy_inhmg_edges}). 
The black curve corresponds to the homogeneous film.
(b) Impact of inhomogeneous $\Lambda$ on $U = -\mu_z B$ calculated using Eq.~(\ref{muz_quadraticF}). 
All calculations in the figures are performed with $\xi/W=0.01$.
   }\label{fig4}
\end{figure}

In principle, we could consider any distribution of $\Lambda(x)$ in our analysis. 
However, for the purpose of illustrating the method of edge-barrier engineering, 
it is useful to consider a simple model that captures its essence. 
We thus adopt a quadratic distribution given by
\begin{eqnarray}
&&\Lambda (x)=\Lambda_0 F(x) , \label{quadratic_Lambda} \\
&&F(x) = \gamma \biggl( \frac{x^2}{W^2} - \frac{1}{4} \biggr) +1 . \label{quadratic_F}
\end{eqnarray}
Here, $\Lambda_0$ represents the Pearl length at the edges of the film, 
and the parameter $\gamma$ controls the variation of $\Lambda$ in the region $-W/2 < x < W/2$. Specifically, for $\gamma < 0$ and $\gamma > 0$, $\Lambda(0) > \Lambda_0$ and $\Lambda(0) < \Lambda_0$, respectively. 
It is worth noting that $\gamma = -\infty$ and $\gamma = 4$ correspond to $\Lambda(0) = \infty$ and $\Lambda(0) = 0$, respectively. 
A homogeneous film is obtained when $\gamma = 0$. 
Figure \ref{fig3} shows the Pearl length distribution $\Lambda(x)$ for different values of $\gamma$.

Using the formulation developed in Sec.\ref{section_formulation}, 
we can calculate the self-energy from Eq.~(\ref{self_energy_inhmg}). 
As shown in Fig.~\ref{fig4} (a), the self-energy of a vortex located at $x=X$ varies depending on the value of $\gamma$. 
When $\gamma>0$, the self-energy of a vortex is larger than that of the homogeneous case ($\gamma=0$), particularly in the middle of the strip due to the shorter Pearl length. 
On the other hand, when $\gamma<0$, the self-energy of a vortex is smaller than that of the homogeneous case.
It is worth noting that the dashed curves, obtained from Eq.~(\ref{self_energy_inhmg_edges}), accurately represent the behavior near the edges of the strip. However, it is interesting to observe that this equation also provides a reasonably good approximation in the middle of the strip, covering a wide range of inhomogeneity parameters.

To obtain the magnetic moment, we can use Eqs.(\ref{MM_inhmg}) and (\ref{inhmg_ML5}). 
Here, $\tilde{\psi}_X(x,0)$ is given by
\begin{eqnarray}
\tilde{\psi}_X(x,0) &=& C_{\pm} \biggl( \tan^{-1}\frac{\sqrt{\gamma}}{\sqrt{4-\gamma}} \mp \tan^{-1}\frac{2x\sqrt{\gamma}}{W\sqrt{4-\gamma}} \biggr) , \\
C_{\pm} &=& W\frac{\tan^{-1}\frac{\sqrt{\gamma}}{\sqrt{4-\gamma}} \pm \tan^{-1}\frac{2X\sqrt{\gamma}}{W\sqrt{4-\gamma}}}{\sqrt{(4-\gamma)\gamma}\tan^{-1}\frac{\sqrt{\gamma}}{\sqrt{4-\gamma}} }
\end{eqnarray}
for $x\ge X$ and $x\le X$, respectively. 
A similar calculation for a film containing an antivortex introduces an additional factor of $(-1)$.
Then, the magnetic moment can be computed as
\begin{eqnarray}
\mu_z(X) = \pm \frac{\phi_0 W^2}{\mu_0 \Lambda_0 \gamma} \ln \frac{4}{4-\gamma + 4\gamma (X/W)^2} , \label{muz_quadraticF}
\end{eqnarray}
for a film including a vortex and an antivortex, respectively. 
Notably, when $\gamma$ approaches $0$, Eq.(\ref{muz_quadraticF}) reduces to the magnetic moment of the homogeneous case given by Eq.(\ref{muz}).
Fig.~\ref{fig4} (b) shows the magnetic potential $U=-\mu_z B$ for a film including a vortex (antivortex) in the magnetic field $B=+2B_{\phi} $ ($B=-2B_{\phi}$). 
For $\gamma>0$, $U$ is more stable than in the homogeneous case.

\begin{figure}[tb]
   \begin{center}
   \includegraphics[height=0.49\linewidth]{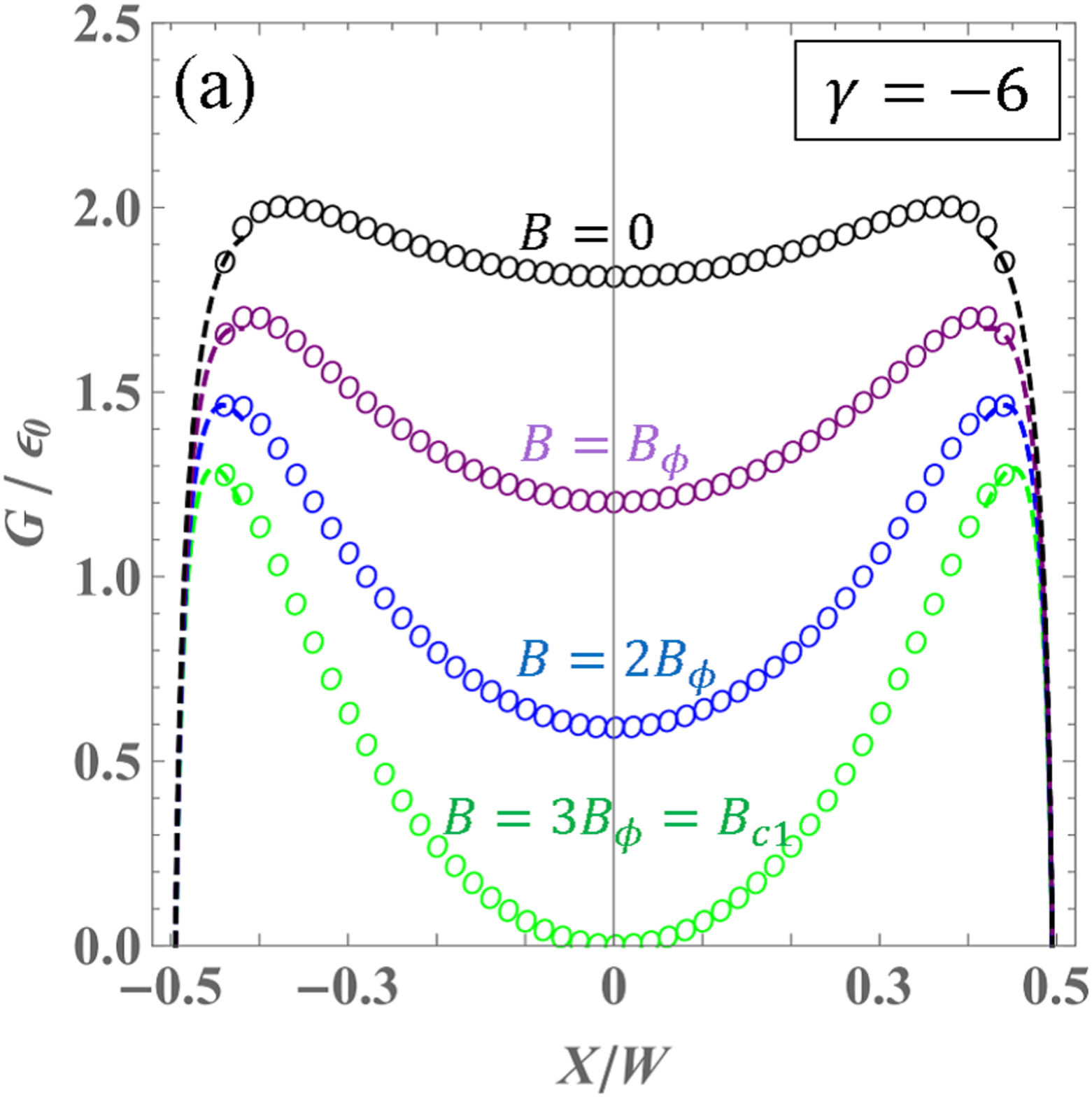}
   \includegraphics[height=0.49\linewidth]{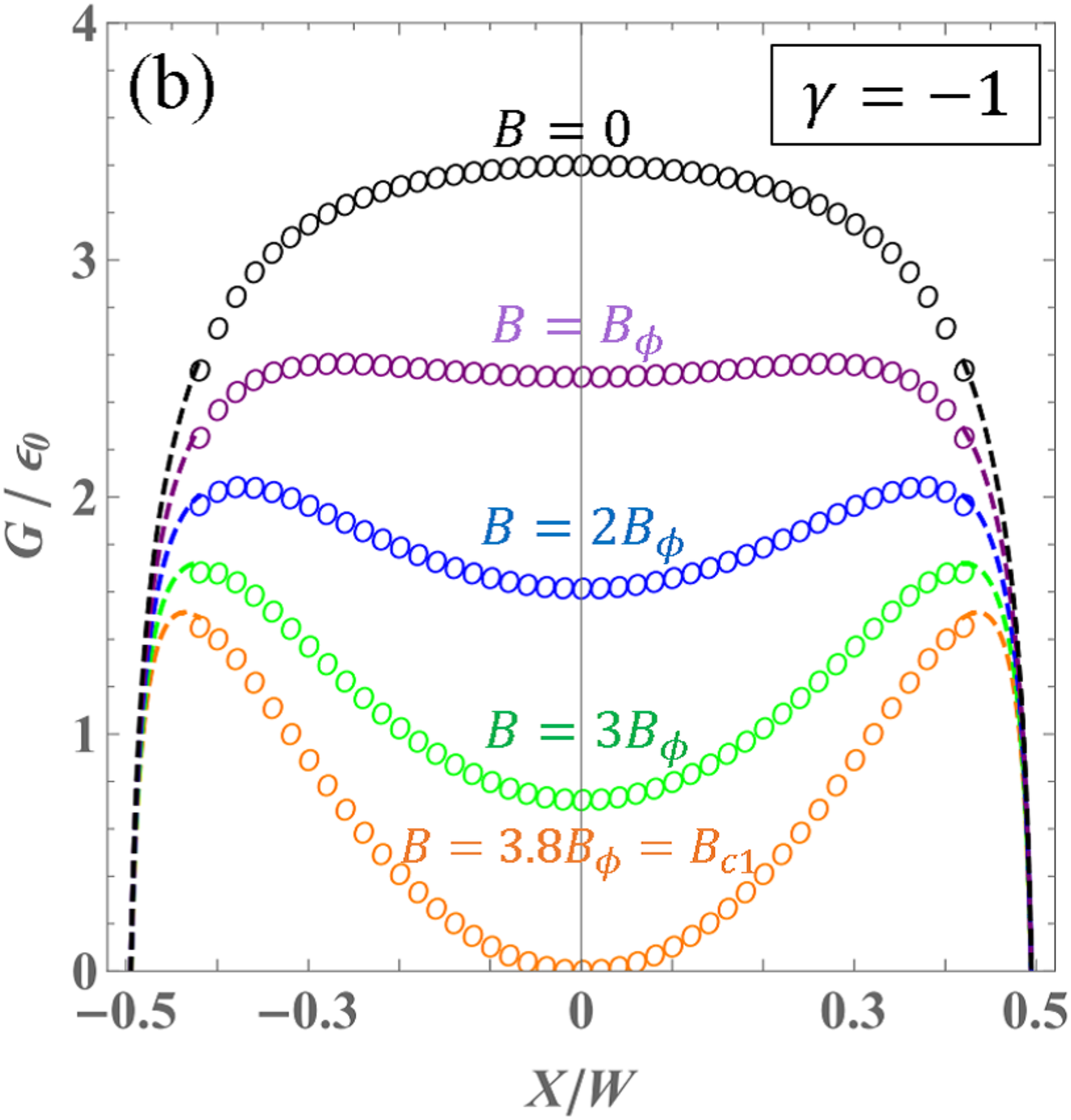}
   \includegraphics[height=0.49\linewidth]{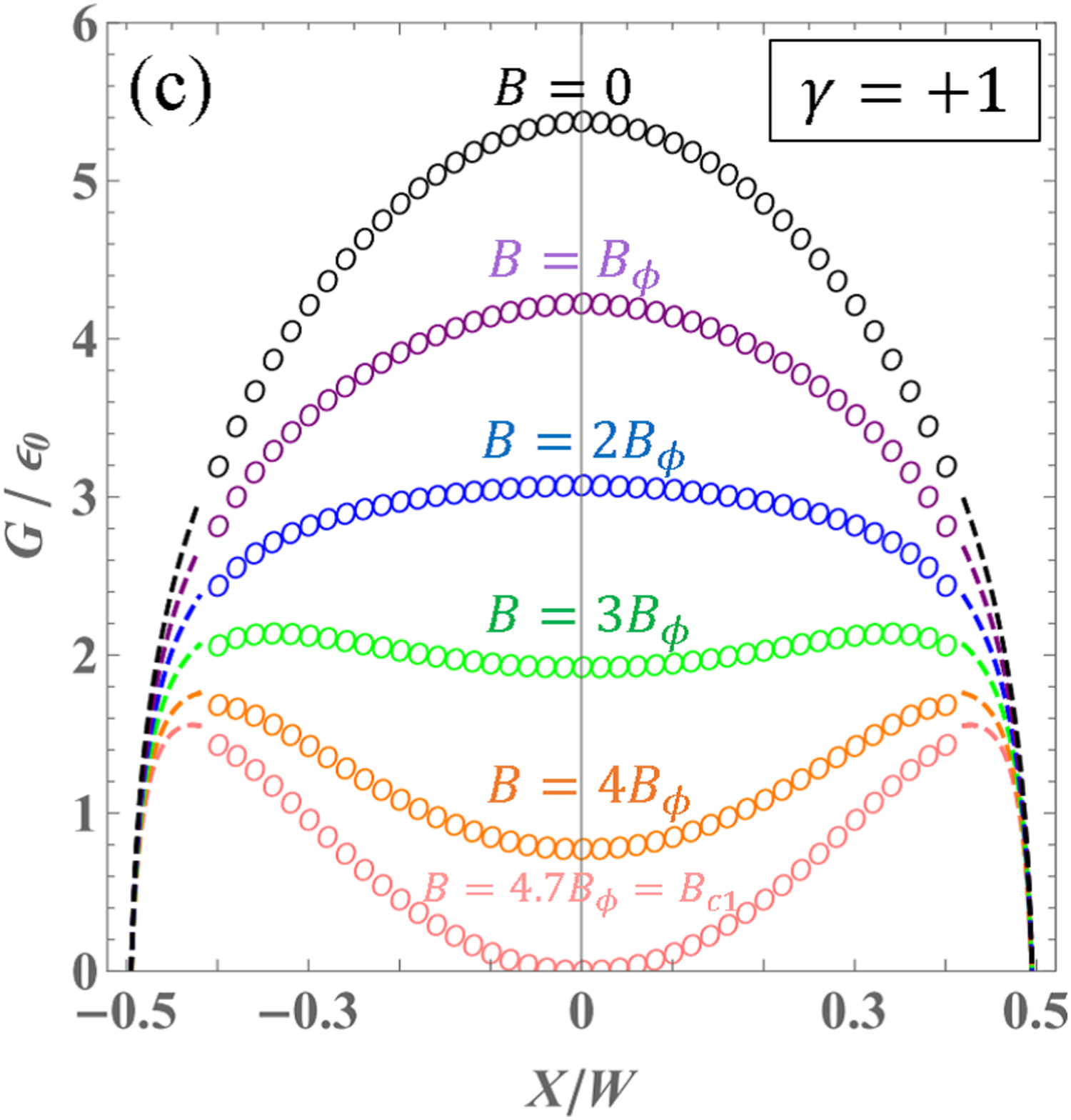}
   \includegraphics[height=0.49\linewidth]{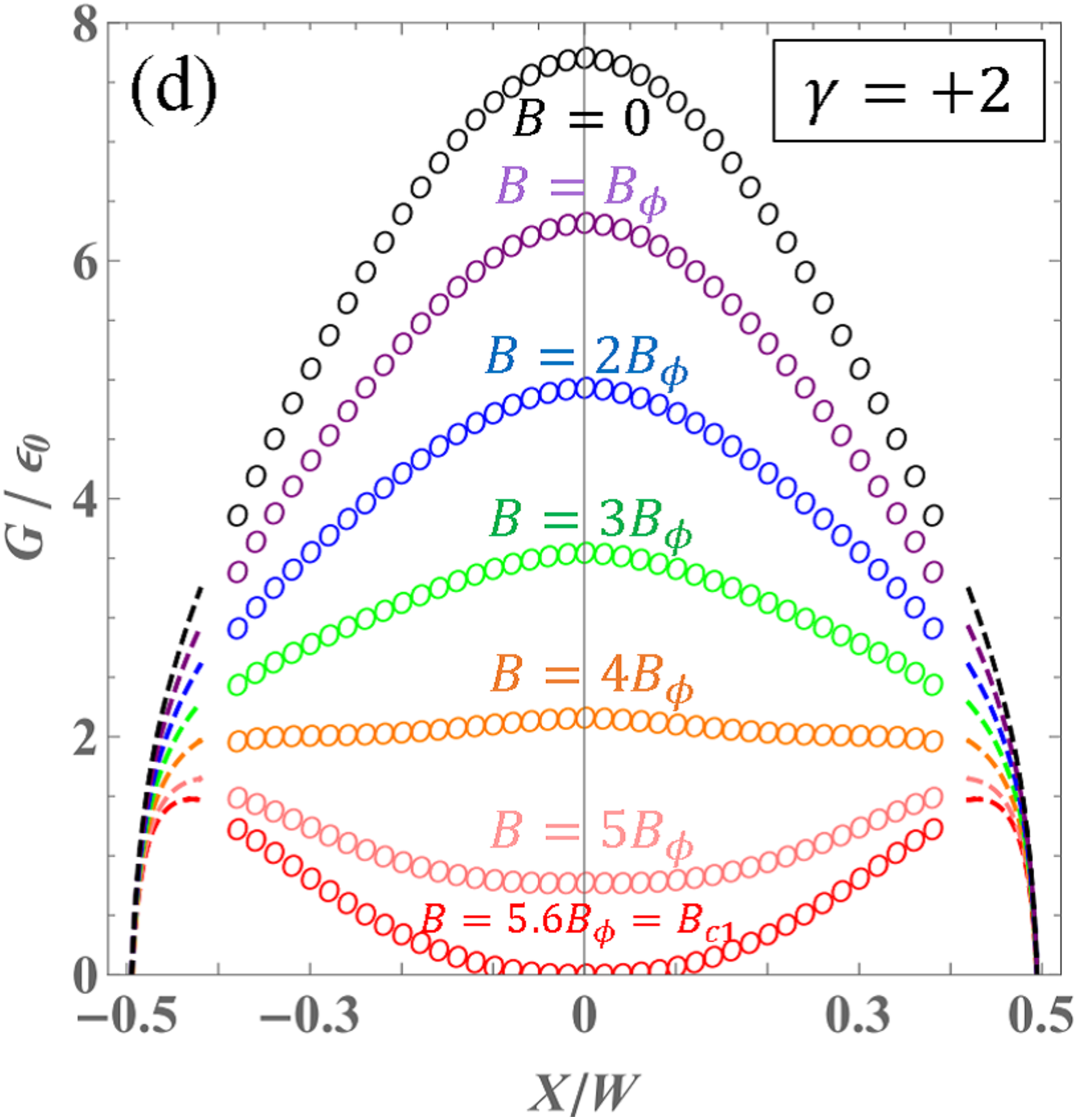}
   \includegraphics[width=0.98\linewidth]{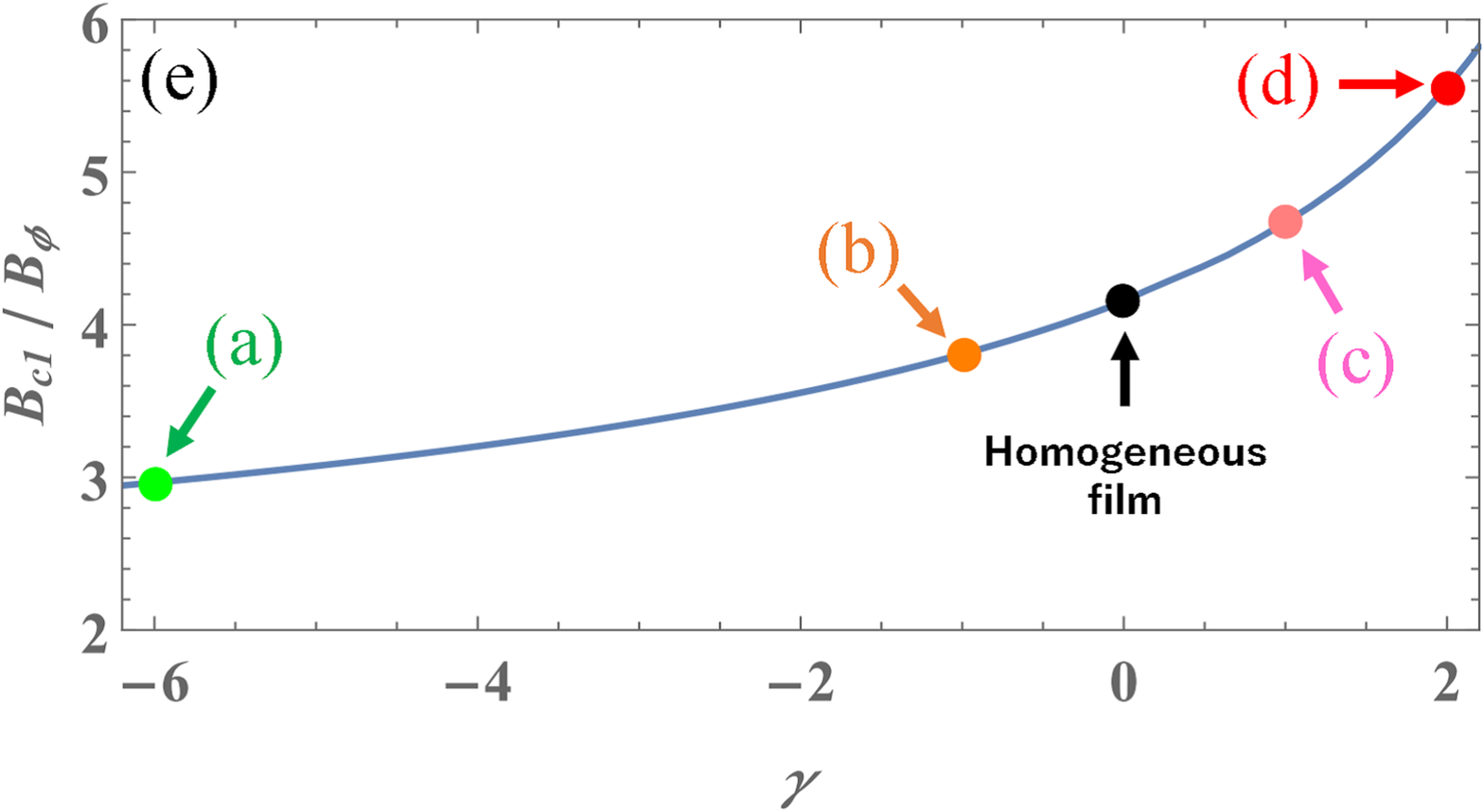}
   \end{center}\vspace{0 cm}
   \caption{
(a-d) Influence of an inhomogeneous $\Lambda(x)$ on the free energy $G=\epsilon - \mu_z B$ in a system with a vortex located at $x=X$. 
The $\Lambda(x)$ distribution is defined by Eqs.~(\ref{quadratic_Lambda}) and (\ref{quadratic_F}) [see also Fig.~\ref{fig3} (a)]. 
The dashed curves near the edges are calculated using Eq.~(\ref{self_energy_inhmg_edges}).
(e) Variation of the critical field $B_{c1}$ as a function of the inhomogeneity parameter $\gamma$. 
The coherence length is fixed at $\xi/W=0.01$ for all calculations in the figures.
   }\label{fig5}
\end{figure}

Fig.\ref{fig5} illustrates the impact of an inhomogeneous $\Lambda(x)$ on the free energy $G=\epsilon-\mu_z B$ of a film containing a vortex in a magnetic field $B\geq 0$. 
The effect of inhomogeneity on $\epsilon$ is dominant compared to $U$, as demonstrated in Fig.\ref{fig4}, and a positive (negative) $\gamma$ stabilizes (destabilizes) the vortex state. 
Notably, when $\gamma=-6$ [Fig~\ref{fig5}(a)], the inhomogeneity stabilizes the vortex state, and the free energy features a metastable minimum even at $B=0$. 
As $\gamma$ increases, the vortex state is destabilized [e.g., compare the black curves in Fig.\ref{fig5}(a-d)]. 
This results in an increase in the critical field $B_{c1}$. 
Fig.\ref{fig5}(e) shows $B_{c1}$ as a function of the inhomogeneity parameter $\gamma$ calculated from Eq.~(\ref{bc1_inhmg}), indicating that an inhomogeneous $\Lambda(x)$ can be employed to engineer $B_{c1}$.

Previously, it was thought that the critical field $B_{c1}$ of a narrow thin-film strip is almost independent of material parameters, and the only practical way to increase $B_{c1}$ was to decrease the strip width. However, we have now discovered that this constraint only applies to homogeneous materials, and we have found another way to increase $B_{c1}$: by designing an inhomogeneous $\Lambda(x)$.

\section{Critical current} \label{section_critical_current}

The method of engineering the free-energy profile is applicable to systems regardless of the presence or absence of a bias current.
This naturally leads to the question: what is the impact of an inhomogeneous Pearl length on the critical current? 
In this section, we will delve into this topic and investigate it further.

\subsection{Critical current in a {\it homogeneous} narrow thin-film strip} \label{section_Ic_hmg}

\begin{figure}[tb]
   \begin{center}
   \includegraphics[width=0.49\linewidth]{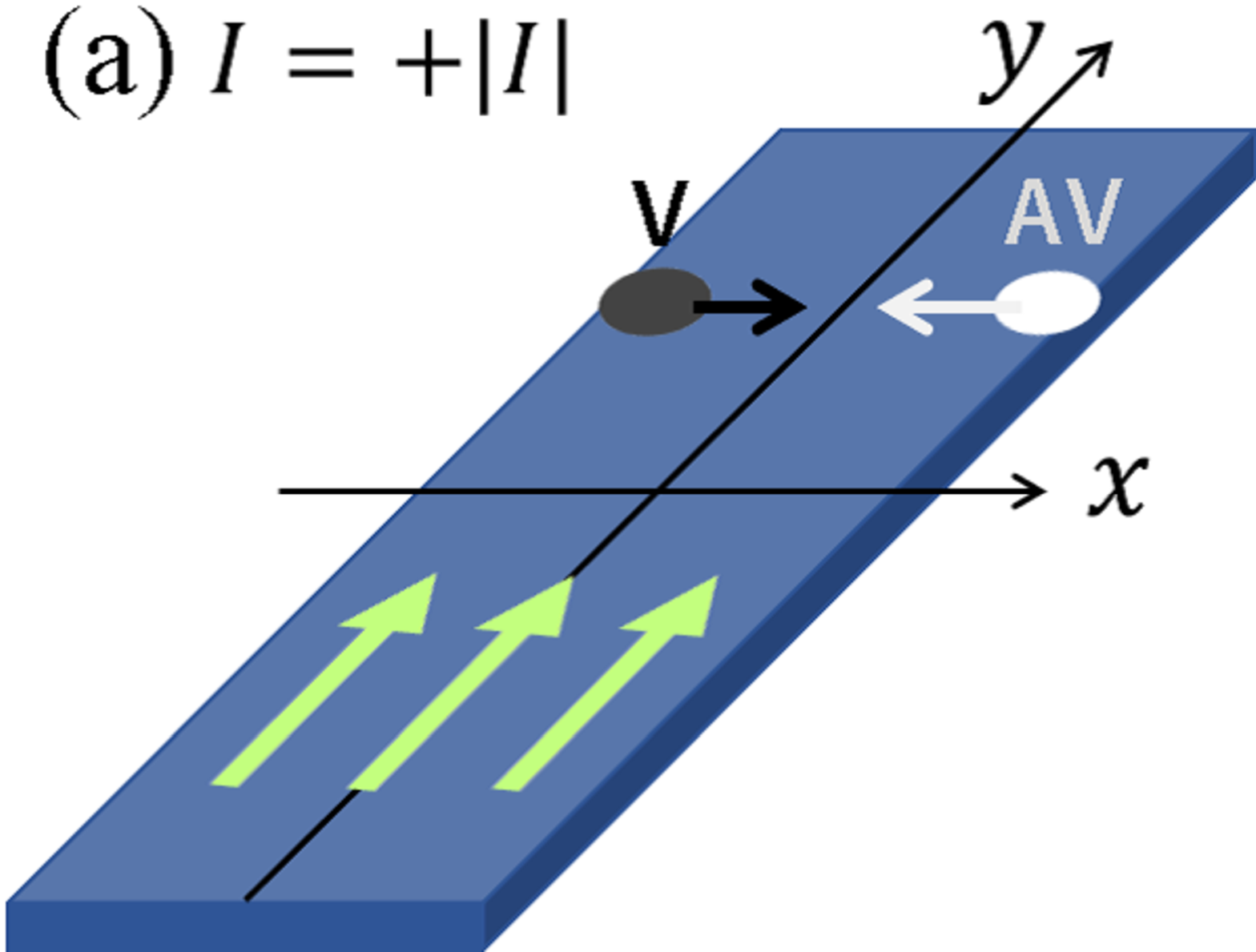}
   \includegraphics[width=0.49\linewidth]{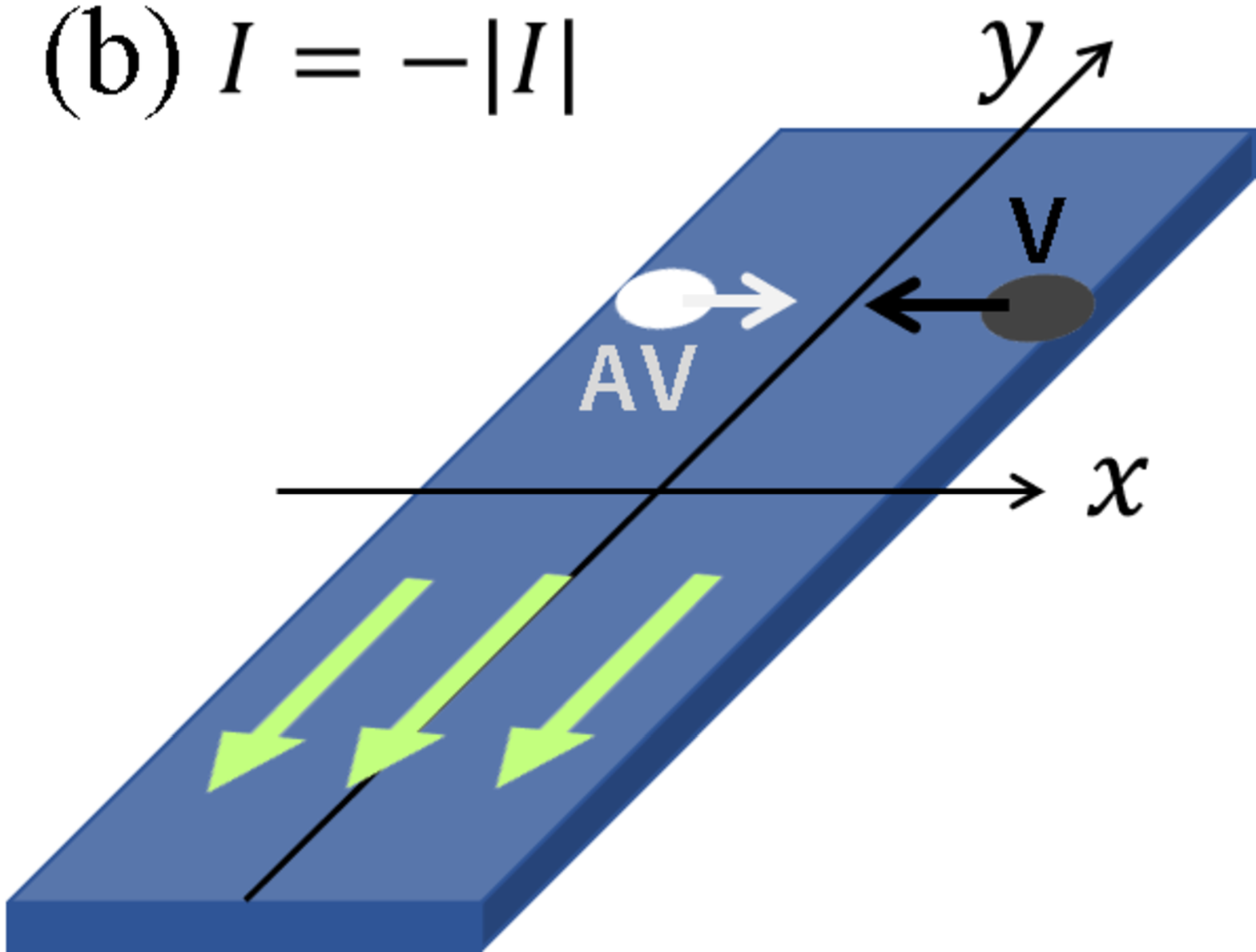}
   \includegraphics[width=0.49\linewidth]{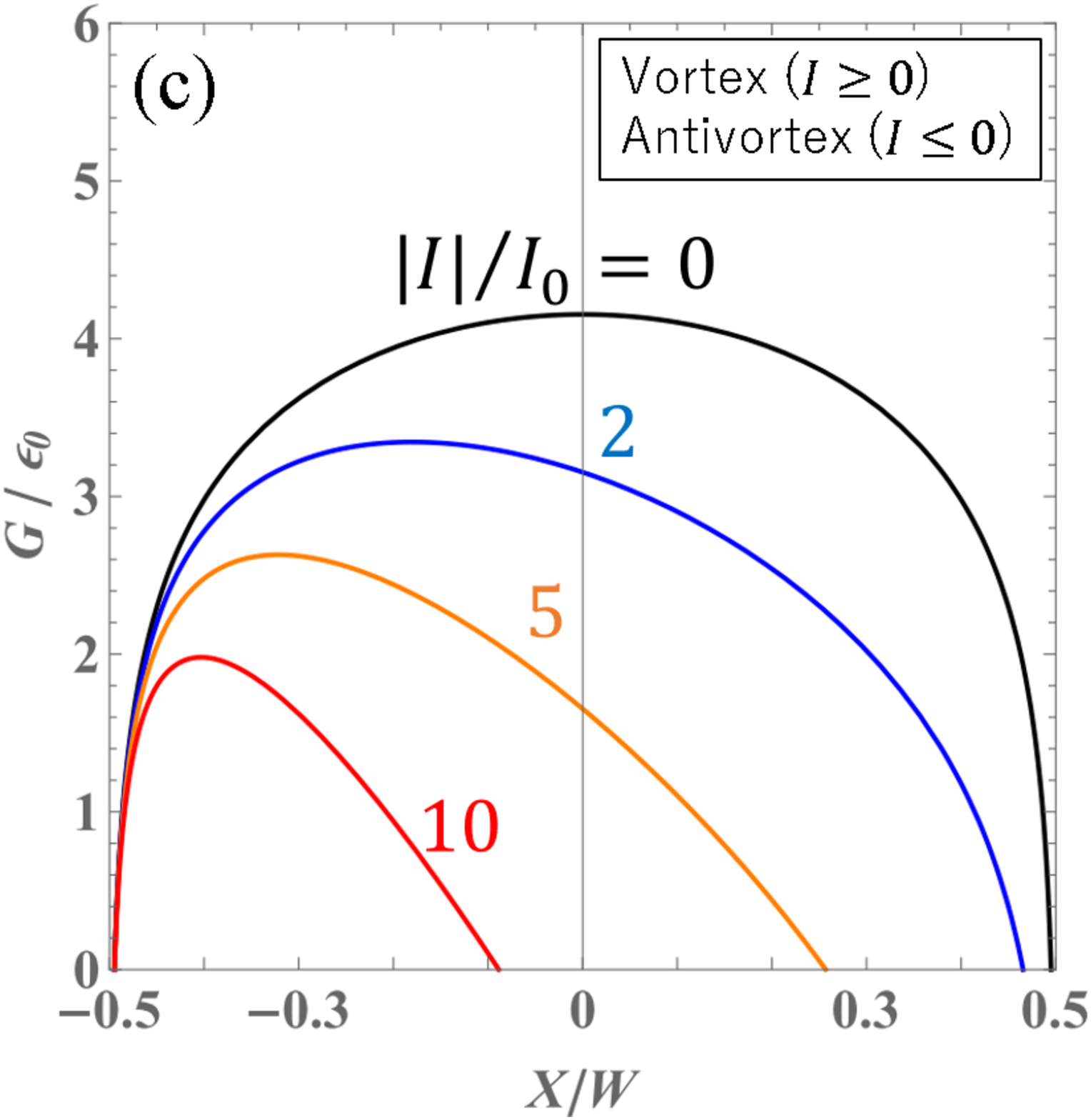}
   \includegraphics[width=0.49\linewidth]{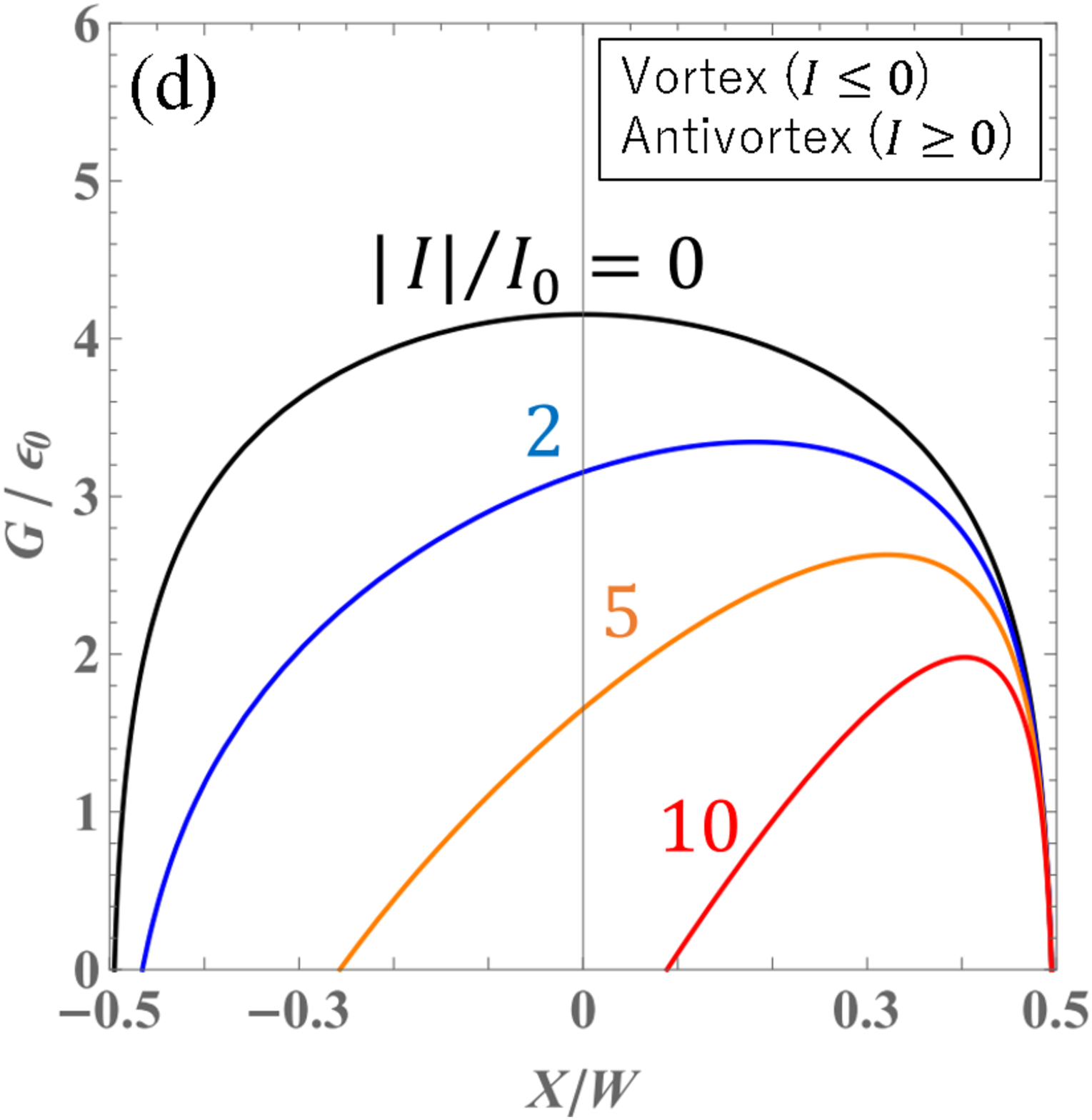}
   \end{center}\vspace{0 cm}
   \caption{
(a, b) Schematic illustrations of the Lorentz forces acting on a vortex and an antivortex in a narrow thin-film strip.
(c, d) Impact of the bias current on the free energy of a homogeneous narrow thin-film strip containing a vortex and an antivortex at $x=X$, with $B=0$. 
   }\label{fig6}
\end{figure}

To begin, we will provide a brief overview of the critical current in a homogeneous narrow thin-film strip.  
Consider a bias current $I$ flowing along the longitudinal direction. 
The sheet bias-current density is given by $J_b = I/W$. 
Then, the free energy of a system containing a vortex or an antivortex located at ${\bf r}_v=(X,0)$ can be expressed as follows:
\begin{eqnarray}
G(X) =  \epsilon(X) - \mu_z(X)B + \Delta G(X) . \label{free_energy_with_bias_current}
\end{eqnarray}
Here, the vortex self-energy $\epsilon(X)$ and the film's magnetic moment $\mu_z(X)$ in a homogeneous film are given by Eqs.~(\ref{self_energy}) and (\ref{muz}), respectively. 
The final term $\Delta G(X)$ represents the work done by the Lorentz force and is expressed as
\begin{eqnarray}
\Delta G(X) = \mp I \phi_0 \biggl[ \frac{X}{W} \pm \frac{ {\rm sgn} (I)}{2} \biggr] \label{deltaG_hmg}
\end{eqnarray}
for a vortex and an antivortex, respectively, 
where ${\rm sgn}$ is the sign function of the current $I$, 
i.e., ${\rm sgn}(I)= \pm 1$ for $I = \pm |I|$, respectively (see Fig.~\ref{fig6}). 
Therefore, we obtain~\cite{2012_Bulaevskii_Graf_Kogan, Maksimova, Clem_Mawatari}
\begin{eqnarray}
&&\frac{G(X)}{\epsilon_0} 
= \ln \biggl( \frac{2W}{\pi\xi} \cos \frac{\pi X}{W} \biggr) \nonumber \\
&&\mp \frac{B}{B_{\phi}} \biggl( 1-4\frac{X^2}{W^2} \biggr) 
\mp \frac{I}{I_0} \biggl[ \frac{X}{W} \pm \frac{ {\rm sgn} (I)}{2} \biggr]
\end{eqnarray}
Here, $I_0$ is introduced as 
\begin{eqnarray}
I_0 = \frac{\phi_0}{2\pi\mu_0\Lambda}.
\end{eqnarray}
To understand the typical magnitude of $I_0$, we can write $I_0= (\xi/W) (\phi_0/4\pi\mu_0 \lambda^2 \xi) (Wd)$. 
Since the second factor is of the order of the depairing current density and the third factor is the cross-sectional area of the strip, 
we find that $I_0$ is smaller than the depairing current $I_d$ by a factor of $\xi/W$.

The vortex-entry current $I_{\rm V}$ (or antivortex-entry current $I_{\rm AV}$) corresponds to the current at which the edge barrier disappears. 
The condition of the disappearance of the edge barrier can be expressed as
\begin{eqnarray}
G(X_p; I_{\rm V, AV}, B)=0 , \label{entry_condition}
\end{eqnarray}
where $X_p$ is the peak position of the edge barrier and satisfies $G'(X_p; I_{\rm V, AV}, B)=0$. 
Solving Eq.~(\ref{entry_condition}), 
we obtain~\cite{2012_Bulaevskii_Graf_Kogan, Maksimova, Tafuri, Clem_Mawatari} (see also Appendix~\ref{derivation})
\begin{eqnarray}
I_{\rm V}^{(\pm)} &=& I_0 \biggl( \frac{2W}{e\xi} - 4\frac{B}{B_{\phi}} \biggr) , \label{I_V_hmg} \\
I_{\rm AV}^{(\pm)} &=& I_0 \biggl( \frac{2W}{e\xi} + 4\frac{B}{B_{\phi}} \biggr) . \label{I_AV_hmg}
\end{eqnarray}
The superscript $(\pm)$ indicates the direction of the current, 
but both $I_{\rm V}$ and $I_{\rm AV}$ are independent of the current direction. 
Here, $e=2.718$ is Napier's constant. 
The critical current is given by 
\begin{eqnarray}
I_c^{(\pm)} = {\rm min} \bigl\{ I_{\rm V}^{(\pm)},  I_{\rm AV}^{(\pm)} \bigr\} 
= 
\begin{cases}
I_{\rm V}^{(\pm)} & (B\ge 0) \\
I_{\rm AV}^{(\pm)} & (B\le 0) 
\end{cases}  \label{Ic_hmg}. 
\end{eqnarray}
Note that $I_c \sim (W/\xi) I_0$, indicating that the magnitude of $I_c$ is approximately equal to the depairing current, $I_c \sim I_d$. 
Additionally, $I_c$ is direction independent.

The linear dependence of $I_c$ on the magnetic field $B$ holds for relatively small values of $B$.
When the magnetic field reaches a critical value $B_{\rm stop}$,
a free energy minimum appears at a point $X_s$ where $G'(X_s; I_c, B_{\rm stop})=0$, 
and penetrating vortices start to reside in the minimum~\cite{Clem_Mawatari, Ilin_Vodo}. 
As these vortices carry a current, they begin to affect the barrier and modify the behavior of $I_c$.
The critical value $B_{\rm stop}$ can be calculated by considering the condition for the existence of a solution $X_s$ that satisfies $G'(X_s; I_c, B_{\rm stop})=0$. 
For a homogeneous film, we obtain
\begin{eqnarray}
B_{\rm stop} = \biggl( s+\frac{1}{4} + \frac{1}{4} \sqrt{8s+1} \biggr) B_{\phi} . \label{Bstop_hmg}
\end{eqnarray}
Here, $s=W/4e\xi$. 
When $\xi/W \ll 1$, Eq.~(\ref{Bstop_hmg}) reduces to $B_{\rm stop}\sim sB_{\phi}=\phi_0/2\pi e W \xi$~\cite{Clem_Mawatari, Ilin_Vodo}.
For example, if $\xi/W=0.005$, we have $s=18$ and $B_{\rm stop}\simeq 22 B_{\phi}$.

\subsection{Critical current in an {\it inhomogeneous} narrow thin-film strip} \label{section_Ic_inhmg}

\subsubsection{Formulation} \label{section_Ic_formulation}

Let us develop the formulation for the critical current in an inhomogeneous narrow thin-film strip. 
When we consider a system with a vortex or an antivortex located at position ${\bf r}_v=(X,0)$ in a film with an inhomogeneous $\Lambda(x)$, the free energy of the system is given by Eq.~(\ref{free_energy_with_bias_current}). 
In an inhomogeneous film, the vortex self-energy $\epsilon(X)$ and the film's magnetic moment $\mu_z(X)$ are given by Eq.(\ref{self_energy_inhmg}) and Eqs.~(\ref{MM_inhmg}) and (\ref{inhmg_ML5}), respectively.
The last term $\Delta G(X)$ corresponds to the work done by the Lorentz force coming from a finite bias current. 
It is important to note that the sheet bias current density $J_b$ is no longer uniform and proportional to the local superfluid density $\Lambda^{-1}(x)$. 
Rather, it is given by
\begin{eqnarray}
J_b(x) =  \frac{I/ \Lambda(x)}{\int_{-W/2}^{W/2} dx [1/\Lambda(x)] } 
=  \frac{I/ F(x)}{\int_{-W/2}^{W/2} dx [1/F(x)] } . \label{inhmg_J}
\end{eqnarray}
It is easy to check that when $\Lambda(x)$ is uniform ($F=1$), 
the sheet bias current density $J_b$ is also uniform and given by $J_b=I/W$. 
Then, we obtain (see also Fig.~\ref{fig6})
\begin{eqnarray}
\Delta G(X) 
&=& \mp  I \phi_0 \frac{\int_{\mp x_{\rm e}}^{X}[1/ F(x)]}{\int_{-W/2}^{W/2} dx [1/F(x)] } ,  \label{delta_G_inhmg}
\end{eqnarray}
for a vortex and an antivortex entry, respectively. 
Here, $x_{\rm e}=(W/2) {\rm sgn}(I)$.

To calculate the critical current, it is necessary to understand the behavior of $G(X)=\epsilon(X)-\mu_z(X) B+\Delta G(X)$ near the edges of the film. 
Fortunately, in this case, Eq.~(\ref{self_energy_inhmg_edges}) provides an analytical expression for $\epsilon(X)$.  
It should be noted that the coherence length $\xi$ depends on $x$, but we only need its value at the edge where the edge barrier disappears.
For certain functions $F(x)$, we can derive analytical expressions for both $\mu_z(X)$ and $\Delta G(X)$,
allowing us to obtain an analytical expression for the critical current $I_c$.
Examples of these expressions will be presented in the following subsections.

\subsubsection{Example: quadratic $\Lambda(x)$} \label{section_critical_current_inhmg}

\begin{figure}[tb]
   \begin{center}
   \includegraphics[width=0.49\linewidth]{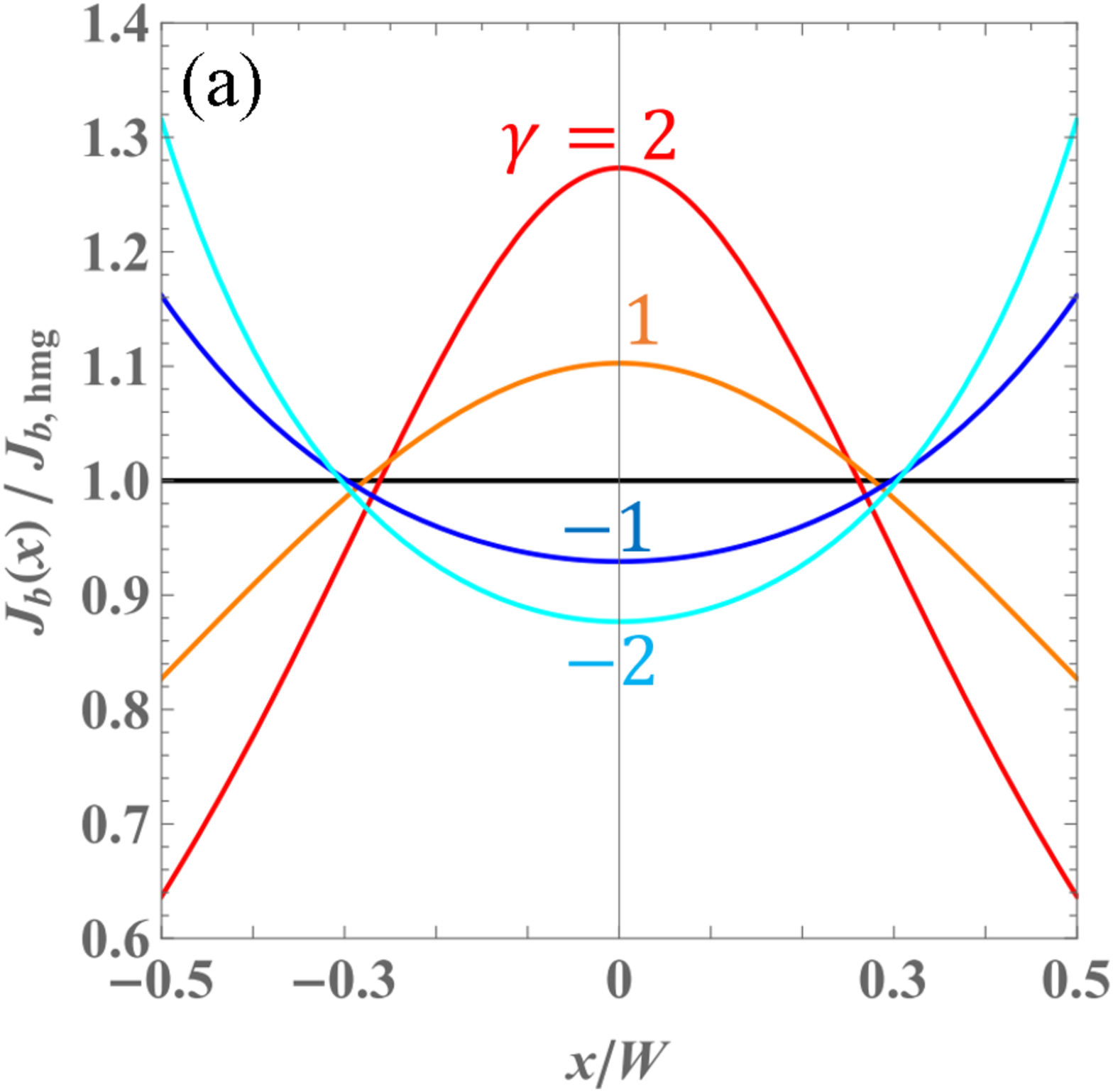}
   \includegraphics[width=0.49\linewidth]{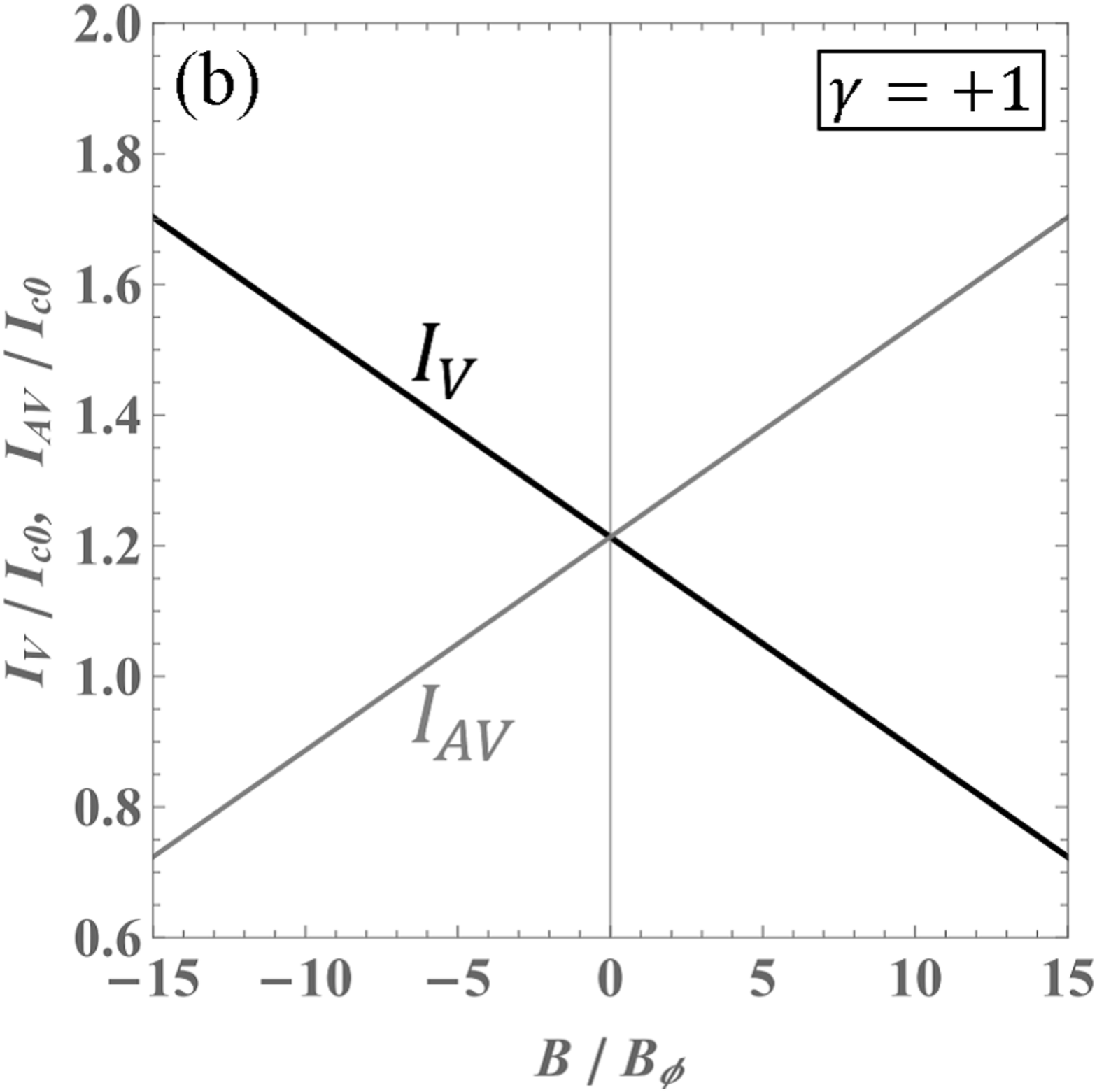}
   \includegraphics[width=0.49\linewidth]{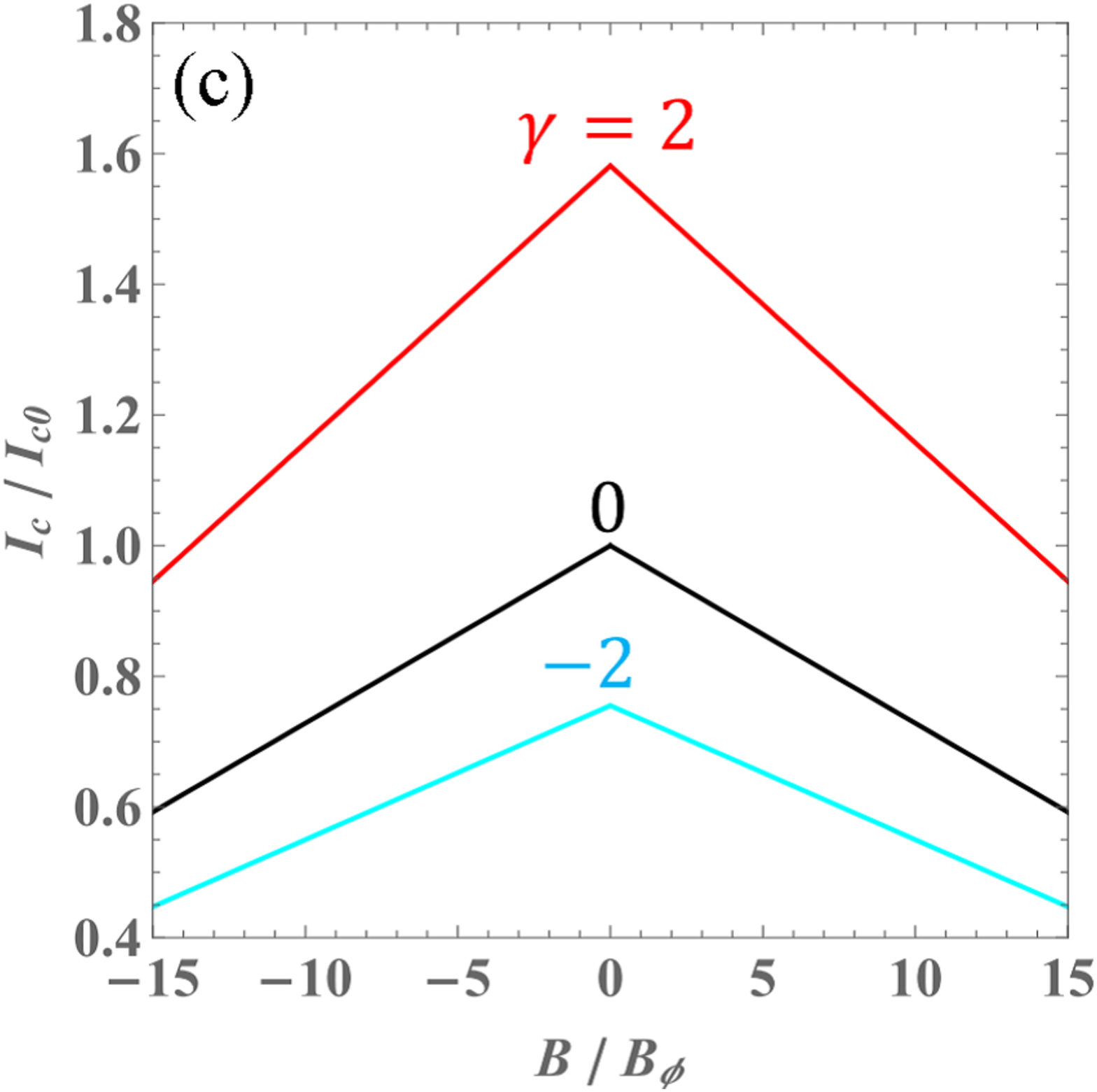}
   \includegraphics[width=0.49\linewidth]{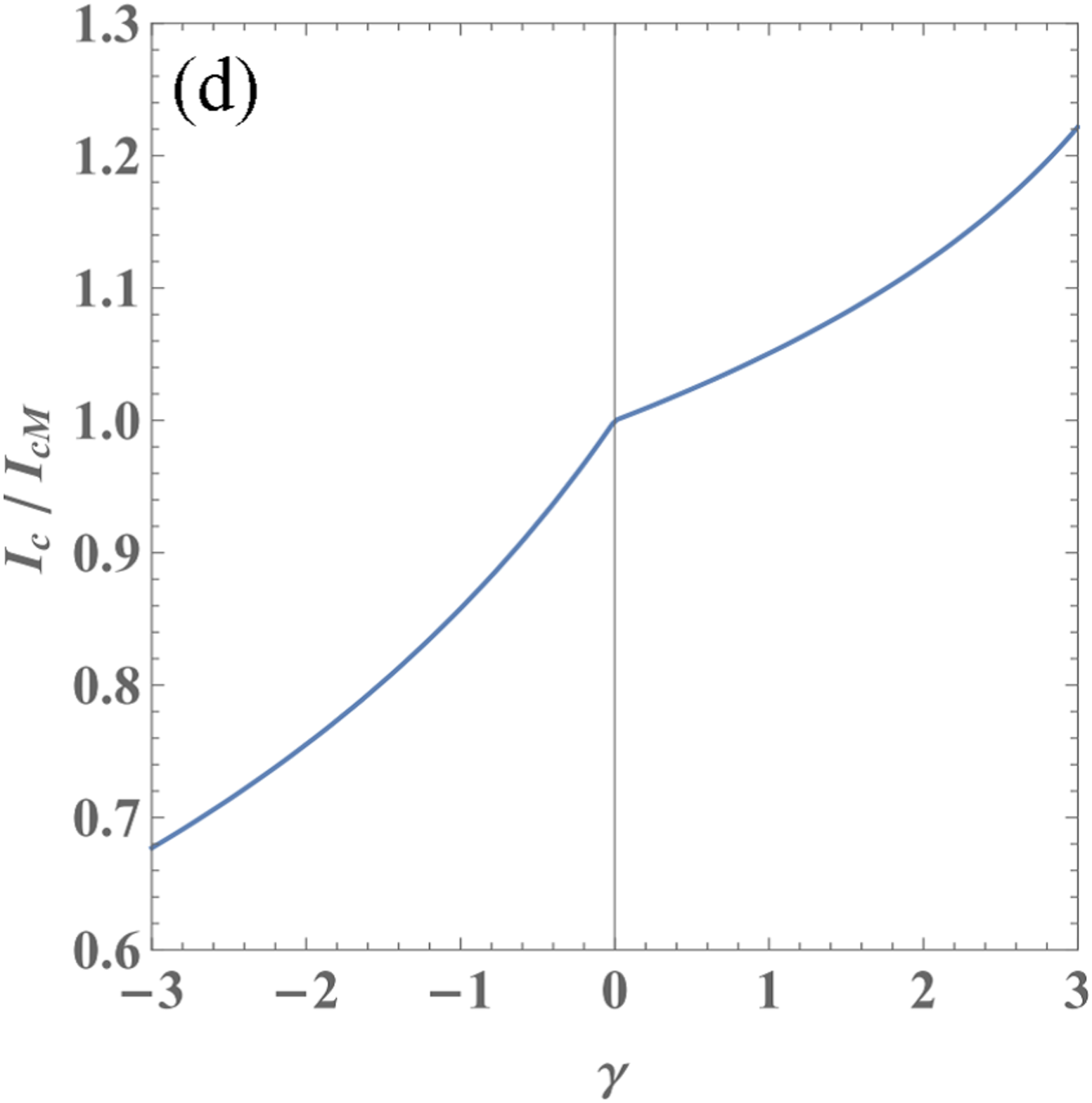}
   \end{center}\vspace{0 cm}
   \caption{
(a) Influence of the inhomogeneity parameter $\gamma$ on the sheet bias-current density $J_b(x)$.
Here, $J_{b, \text{hmg}} = I/W$ [see also Fig.~\ref{fig3} (a)].
(b) Vortex-entry current $I_{\text{V}}$ and antivortex-entry current $I_{\text{AV}}$ as a function of the external magnetic field $B$.
Here, $I_{c0} = (2W/e\xi_e) I_0 = \phi_0 W/e \pi \mu_0 \xi_e \Lambda_0$, where $e$ is Napier's constant and $\xi_e =\xi(\pm W/2)$ is the coherence length at the edges.
(c) Critical currents $I_c = {\rm min} \{ I_{\text{V}}, I_{\text{AV}} \}$ for $\gamma=2$, $0$, $-2$ as a function of $B$.
(d) Comparison of $I_c$ between an inhomogeneous film with $\Lambda(x)$ and a homogeneous film with the shortest Pearl length $\Lambda_{\text{hmg}} = \Lambda_0 \min F(x)$.
Here, $I_{cM}=\phi_0 W/e \pi \mu_0 \xi_{\rm hmg} \Lambda_{\text{hmg}}$. 
For all calculations in the figures, we have assumed $\xi_e/W=0.005$ at the edges.
   }\label{fig7}
\end{figure}

Let us examine the impact of an inhomogeneous distribution of $\Lambda(x)$ on the critical current, considering the same distribution as in Section \ref{section_quadratic} (described by Eqs. (\ref{quadratic_Lambda}) and (\ref{quadratic_F}) and shown in Fig. \ref{fig3}). 
Starting from the expression $G(X)=\epsilon(X)-\mu_z(X)B+\Delta G(X)$, 
we can utilize the self-energy expression given by Eq. (\ref{self_energy_inhmg}) and the magnetic moment $\mu_z$ derived in Section \ref{section_quadratic} [given by Eq. (\ref{muz_quadraticF})]. 
The third term is influenced by the sheet bias-current distribution $J_b(x)$, 
which is determined by the inhomogeneous $\Lambda(x)$. 
Figure \ref{fig7} (a) illustrates $J_b(x)$ for different values of $\gamma$ calculated using Eq. (\ref{inhmg_J}).
Here, $J_{b, hmg}=I/W$. 
For $\gamma>0$ ($\gamma<0$), the superfluid density is higher (lower) in the middle. 
Consequently, the current is more pronounced in the middle (reduced at the edges), 
leading to a suppression (enhancement) of the current at the edges. 
Then, Eq.~(\ref{delta_G_inhmg}) yields 
\begin{eqnarray}
\Delta G(X) = \mp \frac{I}{2 I_0} \biggl( \frac{\tan^{-1}\frac{2X}{W}\sqrt{\frac{\gamma}{4-\gamma}}}{\tan^{-1}\sqrt{\frac{\gamma}{4-\gamma}}} \pm {\rm sgn} (I) \biggr) , 
\end{eqnarray}
for a vortex and an antivortex entry, respectively.
Here, $I_0 = \phi_0/2\pi \mu_0 \Lambda_0$. 
Combining these expressions, we obtain 
\begin{eqnarray}
\frac{G(X)}{\epsilon_0}
&=& \frac{\epsilon(X)}{\epsilon_0}  \mp \frac{B}{B_{\phi}} \frac{4}{\gamma} \ln \frac{4}{4-\gamma + 4\gamma (X/W)^2} \nonumber \\
&\mp& \frac{I}{2 I_0} \biggl( \frac{\tan^{-1}\frac{2X}{W}\sqrt{\frac{\gamma}{4-\gamma}}}{\tan^{-1}\sqrt{\frac{\gamma}{4-\gamma}}} \pm {\rm sgn} (I) \biggr) , \label{G_quadratic}
\end{eqnarray}
for a vortex and an antivortex entry, respectively. 
In the vicinity of the film edges, the self-energy $\epsilon(X)$ can be analytically expressed using Eq.~(\ref{self_energy_inhmg_edges}). 
This analytical expression is sufficient for the calculation of the critical current $I_c$.

The calculations of the vortex-entry current and the antivortex-entry current can be performed straightforwardly using Eqs.~(\ref{self_energy_inhmg_edges}) and (\ref{G_quadratic}) and the condition of the disappearance of the edge barrier given by Eq.~(\ref{entry_condition}). 
We get (see Appendix~\ref{derivation}) 
\begin{eqnarray}
I_{\rm V}^{(\pm)} &=& \frac{4 I_0 \tan^{-1} \sqrt{\frac{\gamma}{4-\gamma}}}{\sqrt{\gamma(4-\gamma)}} \biggl( \frac{2W}{e \xi_e}  - \frac{4 B}{B_{\phi}} \biggr)  , \label{I_V_quadratic} \\
I_{\rm AV}^{(\pm)} &=& \frac{4 I_0 \tan^{-1} \sqrt{\frac{\gamma}{4-\gamma}}}{\sqrt{\gamma(4-\gamma)}} \biggl( \frac{2W}{e \xi_e}  + \frac{4 B}{B_{\phi}} \biggr) . \label{I_AV_quadratic}
\end{eqnarray}
Here, $\xi_e = \xi(\pm W/2)$. 
It is important to note that when performing these calculations, the coherence length at the edges should be used as an input parameter.
Figure \ref{fig7} (b) illustrates the values of $I_{\rm V}=I_{\rm V}^{(\pm)}$ and $I_{\rm AV}=I_{\rm AV}^{(\pm)}$ for $\gamma=1$. 
Due to the system's symmetry, we have $I_V=I_{AV}$ at $B=0$ (see also Fig.~\ref{fig6}).
For $B>0$ ($B<0$), the vortex state (antivortex state) is stabilized, 
resulting in $I_V < I_{AV}$ ($I_{AV} < I_{V}$).

The critical current can be determined using the expression $I_c = I_c^{(\pm)} = {\rm min} \{ I_{\rm V}^{(\pm)}, I_{\rm AV}^{(\pm)} \}$. 
Figure~\ref{fig7} (c) illustrates the variation of the critical current $I_c$ for different values of $\gamma$.
It should be noted that, as discussed in Section~\ref{section_Ic_hmg}, the linear dependence of $I_c$ on $B$ ceases to hold when $B \ge B_{\rm stop}$. 
In an inhomogeneous film, $B_{\rm stop}$ is no longer given by Eq.~(\ref{Bstop_hmg}) and depends on the inhomogeneity parameter $\gamma$. 
However, as shown in Appendix~\ref{appendix_Bstop}, we typically have $B_{\rm stop}(\gamma) \gtrsim 20$ for $\gamma \gtrsim -3$. Within the range of the plotted data, the linearity of the critical current with respect to $B$ continues to hold.

Now, let us investigate whether an inhomogeneous film can enhance the critical current. 
We assume that the inhomogeneity arises from a nonuniform impurity concentration. 
In a homogeneous film with a constant value of $\Lambda_{\rm hmg}$, the critical current is given by the formula $I_{c}=\phi_0 W/e\pi \mu_0 \xi_{\rm hmg} \Lambda_{\rm hmg}$ at $B=0$. 
Unlike the critical field $B_{c1}$, the critical current of a homogeneous film can be improved by simply reducing the product $\xi_{\rm hmg} \Lambda_{\rm hmg}$. 
This suggests that instead of creating an inhomogeneous film with varying $\Lambda(x)$, we can achieve an increase in the critical current by using a homogeneous film with a lower impurity concentration.
Assuming the capability to fabricate an inhomogeneous film with varying $\Lambda(x)$, it is also assumed that we have the technology to create a corresponding homogeneous film with $\Lambda_{\rm hmg} = \min \Lambda(x) = \Lambda_0 \min F(x)$.
Here, $\min F(x)=1$ for $\gamma\le 0$ and $\min F(x)=F(0)=1-\gamma/4$ for $\gamma\ge 0$ (see Fig.~\ref{fig3}). 
The coherence length of such a homogeneous film is given by $\xi_{\rm hmg} = \xi_e/\sqrt{\min F(x)}$. 
In this context, we are interested in comparing the critical current $I_c$ of the inhomogeneous film with varying $\Lambda(x)$ to the critical current $I_{cM}$ of the homogeneous film with a constant $\Lambda_{\rm hmg}$. 
Figure~\ref{fig7} (d) illustrates this comparison, clearly demonstrating that an inhomogeneous film with $\gamma>0$ (lower impurity concentration in the middle of the strip) consistently exhibits a larger critical current compared to the homogeneous film. 
The enhancement in critical current is attributed to the current-suppression effect at the edges [see Fig.~\ref{fig7} (a)]. This mechanism is analogous to the superheating field enhancement structure studied in superconducting resonator technology for particle accelerators~\cite{2006_Gurevich, 2014_Kubo, 2015_Gurevich, 2017_Liarte_SUST, 2017_Kubo_SUST, 2019_Sauls, 2021_Kubo}.  
These results indicate that, for the purpose of enhancing the critical current, utilizing an inhomogeneous film with an optimally designed $\Lambda(x)$ is preferable over a homogeneous film.

\subsubsection{Example: Left-right asymmetric linear $\Lambda(x)$ and superconducting diode effect} \label{section_diode}

\begin{figure}[tb]
   \begin{center}
   \includegraphics[width=0.49\linewidth]{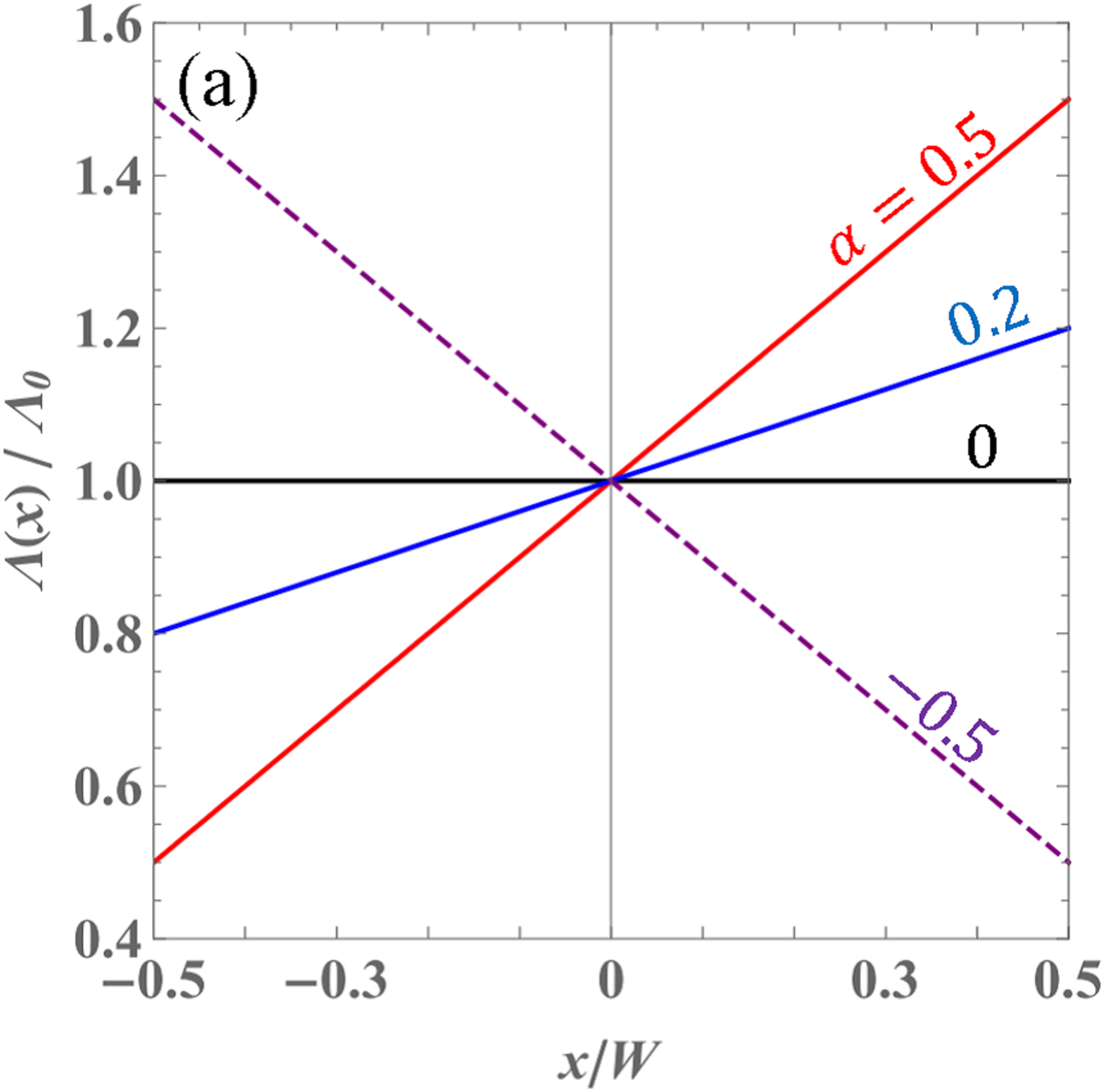}
   \includegraphics[width=0.49\linewidth]{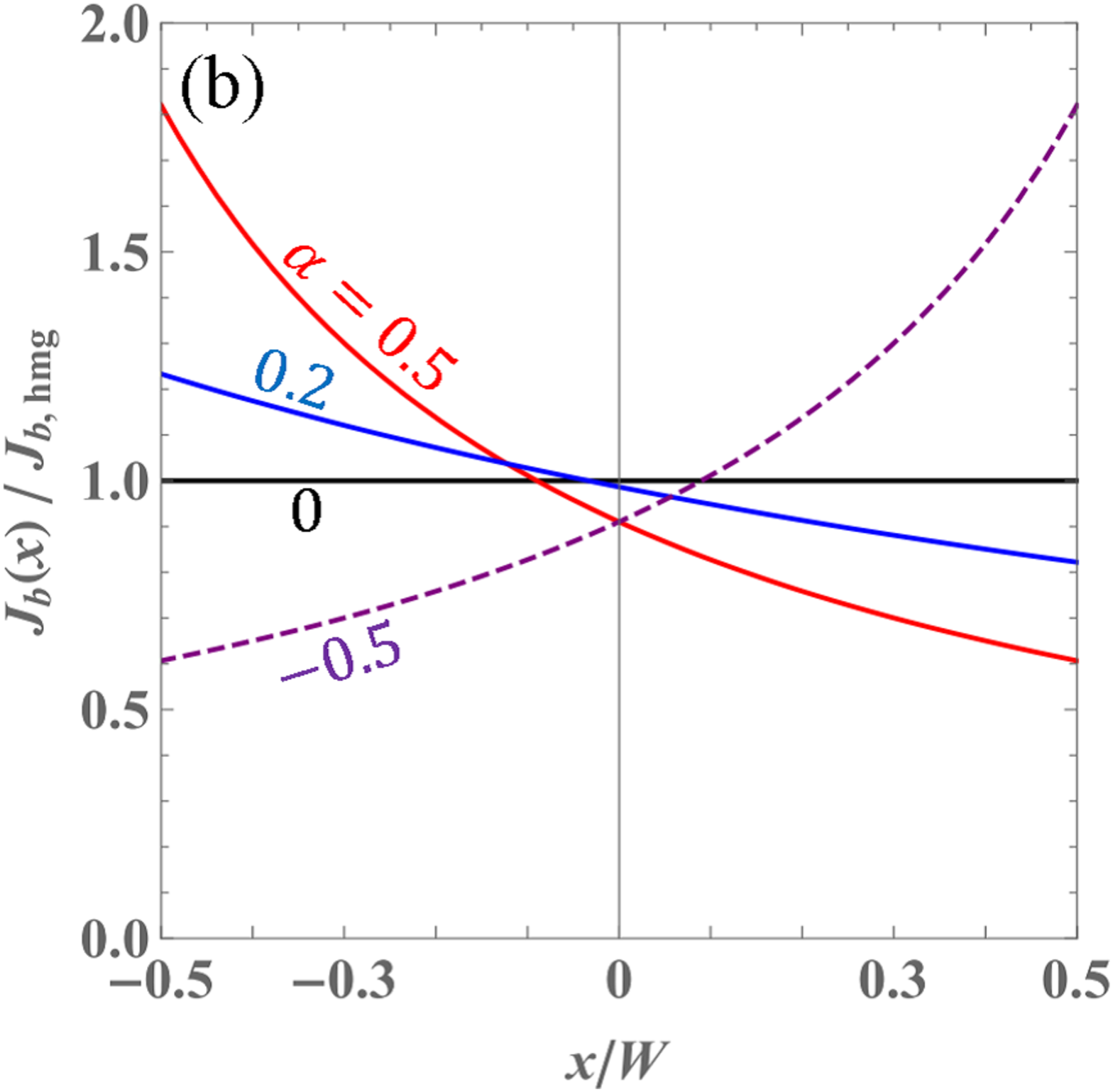}
   \end{center}\vspace{0 cm}
   \caption{
(a) Examples of left-right asymmetric Pearl length distribution $\Lambda(x)$ given by Eqs.~(\ref{linear_Lambda}) and (\ref{linear_F}) for different $\alpha$ values. 
(b) Sheet bias-current distributions $J_b(x)$ for different $\alpha$ values. 
Here, $J_{hmg} = I/W$ is the sheet bias-current for a homogeneous strip. . 
   }\label{fig8}
\end{figure}

As we saw in Section~\ref{section_Ic_hmg}, the critical current of a homogeneous film does not depend on the direction of the current. 
This remains true even for an inhomogeneous film with the quadratic $\Lambda(x)$ given by Eqs.~(\ref{quadratic_Lambda}) and (\ref{quadratic_F}), as discussed in Section~\ref{section_critical_current_inhmg}. 
This fact is summarized by the vanishing of the superconducting diode quality parameter $\eta$, 
which measures the nonreciprocity of the critical current, 
\begin{eqnarray}
\eta = \frac{I_c^{+}-I_c^{-}}{I_c^{+} + I_c^{-}} . \label{eta}
\end{eqnarray}
It is known that $\eta$ is nonzero only when inversion and time-reversal symmetries are broken.

To make $\eta$ nonzero in our system, a combination of left-right asymmetric edge-barriers and an external magnetic field is necessary. 
While any form of $\Lambda(x)$ can be used as long as it is left-right asymmetric, 
we consider the following simplest form in this article [see Fig.~\ref{fig8} (a)]:
\begin{eqnarray}
&&\Lambda(x) = \Lambda_0 F(x), \label{linear_Lambda} \\
&&F(x) = 2\alpha \frac{x}{W} + 1 . \label{linear_F}
\end{eqnarray}
Here, $\Lambda_0=[\Lambda(W/2)+\Lambda(-W/2)]/2$, 
$\alpha = [\Lambda(W/2) - \Lambda(-W/2)]/[\Lambda(W/2) + \Lambda(-W/2)]$.  
For instance, $\alpha=0.5$ corresponds to $\Lambda(W/2)/\Lambda(-W/2)=3$. 
The inhomogeneous $\Lambda(x)$ distribution leads to a non-uniform sheet bias-current density $J_b$, which can be obtained using Eq.~(\ref{inhmg_J}). 
Fig.~\ref{fig8}(b) displays the profiles of $J_b(x)$ for different values of $\alpha$.

Let us analyze the free energy $G=\epsilon-\mu_z B + \Delta G$. 
The magnetic potential term is computed using Eqs.~(\ref{MM_inhmg}) and (\ref{inhmg_ML5}). 
The solution of Eq.(\ref{inhmg_ML5}) is given by
\begin{eqnarray}
\tilde{\psi}_X(x,0)
= \frac{-W}{4\alpha \tanh^{-1}\alpha} \ln \frac{1\mp \alpha}{1+2\alpha \frac{X}{W}} \ln \frac{1\pm \alpha}{1+2\alpha \frac{x}{W}}
\end{eqnarray}
for $x\ge X$ and $x\le X$, respectively. 
A similar calculation for a film containing an antivortex introduces an additional factor of $(-1)$.
Using Eq.~(\ref{MM_inhmg}), we get the magnetic moment
\begin{eqnarray}
\mu_z(X)&=&\frac{\pm \phi_0 W^2}{2\mu_0 \Lambda_0 \alpha \tanh^{-1}\alpha} 
\biggl[ -\ln \sqrt{1-\alpha^2}  \nonumber \\
&& - \frac{X}{W} \ln\frac{1+\alpha}{1-\alpha} + \ln \biggl( 1+ 2\alpha \frac{X}{W} \biggr) \biggr].
\end{eqnarray}
for a film containing a vortex and antivortex, respectively. 
The work done by the Lorentz force can be calculated from Eq.~(\ref{delta_G_inhmg}), 
and we get
\begin{eqnarray}
\Delta G(X) = \mp I \phi_0  \frac{\ln[(1+2\alpha X/W)/\{ 1\mp \alpha {\rm sgn}(I) \} ]}{\ln [(1+\alpha)/(1-\alpha)]}
\end{eqnarray}
Combining these terms, 
we obtain 
\begin{eqnarray}
&&\frac{G(X)}{\epsilon_0} 
= \frac{\epsilon(X)}{\epsilon_0}  \nonumber \\
&&\mp \frac{B}{B_{\phi}}\frac{2}{\alpha \tanh^{-1}\alpha} 
\biggl[ -\ln \sqrt{1-\alpha^2}  \nonumber \\
&& - \frac{X}{W} \ln\frac{1+\alpha}{1-\alpha} + \ln \biggl( 1+ 2\alpha \frac{X}{W} \biggr) \biggr] \nonumber \\
&&\mp \frac{I}{I_0} \frac{\ln[(1+2\alpha X/W)/\{ 1\mp \alpha {\rm sgn}(I) \} ]}{\ln [(1+\alpha)/(1-\alpha)]}. \label{G_linear}
\end{eqnarray}
for a vortex and antivortex, respectively. 
The vortex self-energy term $\epsilon(X)$ is calculated using Eq.~(\ref{self_energy_inhmg}) for the region $-W/2<X<W/2$, 
while the analytical expression provided by Eq.~(\ref{self_energy_inhmg_edges}) is applicable for the edges.

The vortex-entry current $I_{\rm V}^{(\pm)}$ and the antivortex-entry current $I_{\rm AV}^{(\pm)}$ can be determined by calculating the current at which the edge barrier disappears, as discussed in Section~\ref{section_Ic_hmg} and \ref{section_critical_current_inhmg}, 
using Eq.~(\ref{entry_condition}). 
We find (see Appendix~\ref{derivation})
\begin{eqnarray}
\frac{I_{\rm V}^{(\pm)}}{I_0} &=& \frac{\ln[(1+\alpha)/(1-\alpha)]}{2\alpha} \biggl[ \frac{2W}{e\xi_{L, R}}  \nonumber \\
&&\mp  \frac{2 (B/B_{\phi})}{\alpha \tanh^{-1}\alpha} \biggl\{ 2 \alpha - (1\mp \alpha)\ln\frac{1+\alpha}{1-\alpha}  \biggr\} \biggr] , \label{I_V_linear} \\
\frac{I_{\rm AV}^{(\pm)}}{I_0} &=& \frac{\ln[(1+\alpha)/(1-\alpha)]}{2\alpha} \biggl[ \frac{2W}{e\xi_{R, L}}  \nonumber \\
&&\mp  \frac{2 (B/B_{\phi})}{\alpha \tanh^{-1}\alpha} \biggl\{ 2 \alpha - (1\pm \alpha)\ln\frac{1+\alpha}{1-\alpha}  \biggr\} \biggr] .\label{I_AV_linear} 
\end{eqnarray}
Here, $\xi_L=\xi(-W/2)$ and $\xi_R=\xi(+W/2)$. 
It is important to note that in the calculations of $I_{\rm V, AV}^{(\pm)}$, the coherence lengths at the edges are required. 
In this section, we consider the inhomogeneity to be a result of variations in impurity concentration. To determine the coherence length at the edges, we calculate $\xi_L=\xi_M/\sqrt{F(-W/2)}=\xi_M/\sqrt{1-\alpha}$ and $\xi_R=\xi_M/\sqrt{F(W/2)}=\xi_M/\sqrt{1+\alpha}$, where $\xi_M=\xi|_{x=0}$ represents the coherence length at the middle of the strip.

\begin{figure}[tb]
   \begin{center}
   \includegraphics[width=0.49\linewidth]{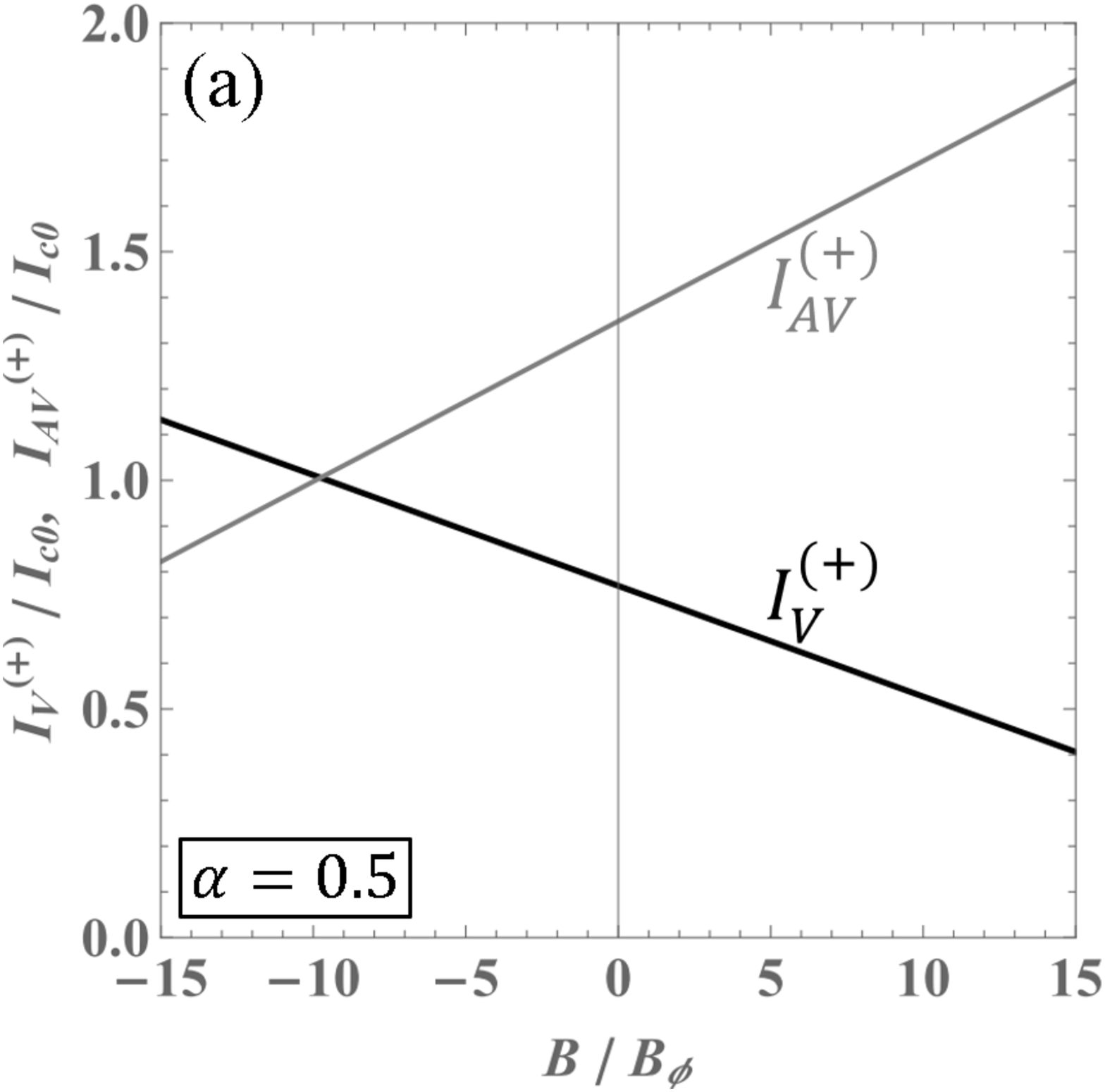}
   \includegraphics[width=0.49\linewidth]{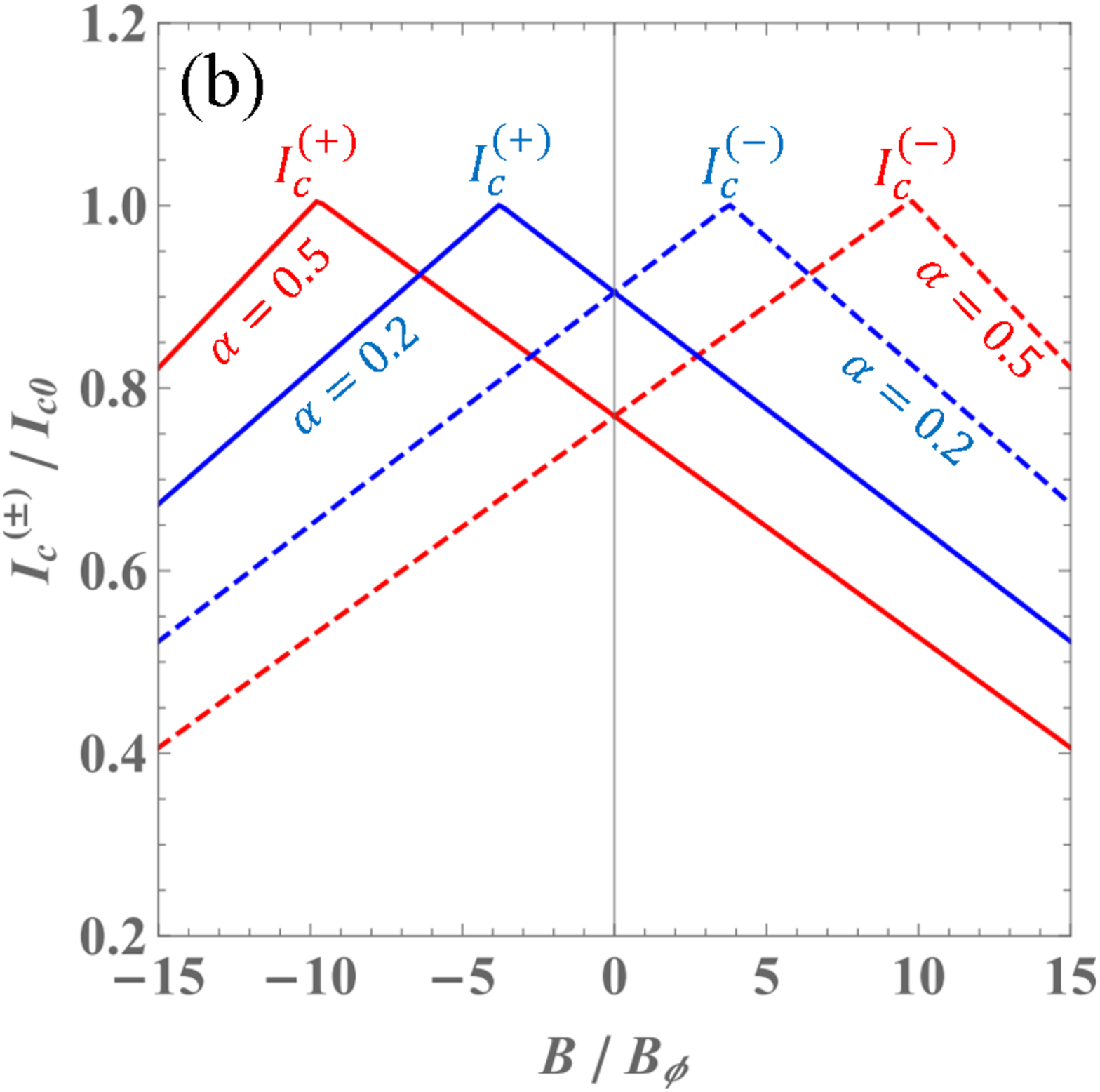}
   \includegraphics[width=0.98\linewidth]{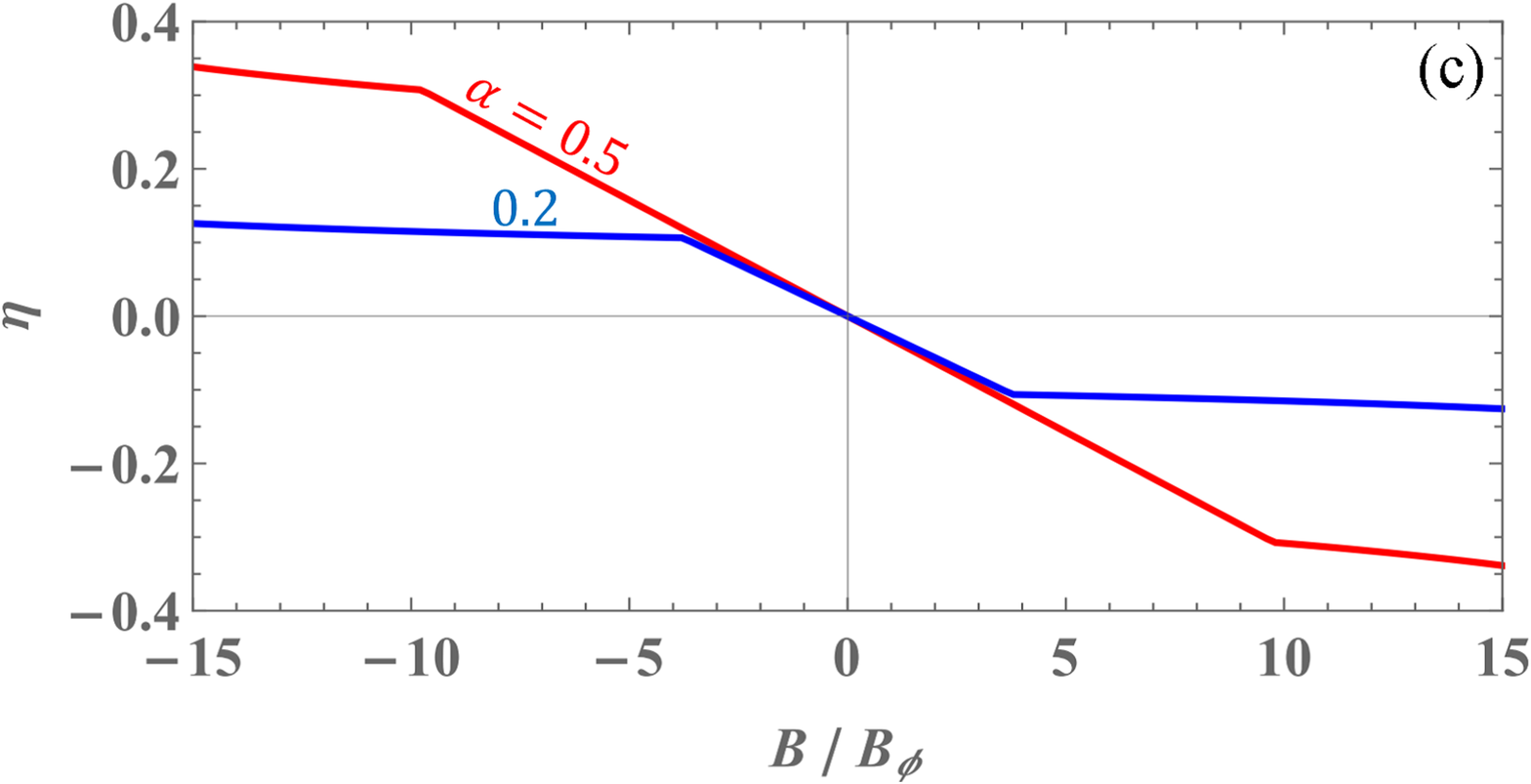}
   \end{center}\vspace{0 cm}
   \caption{
(a) Examples of $I_{\rm V}^{(+)}$ and $I_{\rm AV}^{(+)}$, normalized by $I_{c0} = (2W/e\xi) I_0 = \phi_0 W/e \pi \mu_0 \xi_M \Lambda_0$. 
Here, $\xi_M=\xi|_{x=0}$. 
(b) The critical currents $I_c^{(+)}$ (solid line) and $I_c^{(-)}$ (dashed line) as functions of the external field $B$ for inhomogeneity parameters $\alpha=0.5$ (red) and $0.2$ (blue). 
(c) The superconducting diode quality parameter $\eta$ as functions of $B$ for different $\alpha$.
For all calculations in the figures, we have used $\xi_M/W=0.005$.
   }\label{fig9}
\end{figure}

Figure \ref{fig9} (a) presents examples of $I_{\rm V}^{(+)}$ and $I_{\rm AV}^{(+)}$. 
At $B=0$, we find that $I_{\rm V}^{(+)}<I_{\rm AV}^{(+)}$, which is in contrast to the left-right symmetric cases discussed in the previous subsections where $I_{\rm V}^{(+)}=I_{\rm AV}^{(+)}$ at $B=0$. 
One might expect the left edge to have a higher energy barrier compared to the right edge due to the smaller $\Lambda$ for $\alpha>0$, leading to the expectation of $I_{\rm V}^{(+)}>I_{\rm AV}^{(+)}$.
However, the strong Lorentz force induced by the enhanced sheet bias-current at the left edge [see Fig.~\ref{fig8} (b)], resulting from the left-right asymmetric $\Lambda(x)$ distribution, overcomes the barrier and pushes the vortex into the interior of the film. 
Consequently, for $\alpha>0$ at $B=0$, we observe $I_{\rm V}^{(+)}<I_{\rm AV}^{(+)}$.
As $B$ increases, the vortex state becomes more stable (while the antivortex state becomes less stable), resulting in a decrease in the vortex-entry current $I_{\rm V}$ and an increase in the antivortex-entry current $I_{\rm AV}$.

The critical current can be determined straightforwardly by evaluating $I_c^{(\pm)} = \min \{ I_{\rm V}^{(\pm)}, I_{\rm AV}^{(\pm)} \}$. 
Figure \ref{fig9} (b) shows the critical currents $I_c^{(+)}$ and $I_c^{(-)}$ for various values of $\alpha$. 
It can be verified that $I_c^{(+)}$ for $\alpha = 0.5$ is derived from $I_{\rm V}^{(+)}$ and $I_{\rm AV}^{(+)}$ in Fig.~\ref{fig9} (a). 
The dependence of the maximum values of $I_c^{(\pm)}$ on the inhomogeneity parameter $\alpha$ is found to be remarkably weak. This weak dependence can be attributed to the accidental cancellation between the influences of $\alpha$ and $\xi_{L,R}(\alpha)$ on $I_c$. 
Fig.~\ref{fig9} (c) presents the superconducting diode quality parameter $\eta$ as a function of $B$ for different values of $\alpha$. 
Note that, as demonstrated in Appendix~\ref{appendix_Bstop}, the critical value $B_{\rm stop}(\alpha)$ is greater than 15 for $|\alpha| \lesssim 0.5$. 
Therefore, within the range of the presented data, the linear relationship between the critical current and magnetic field, $B$, remains valid.

\section{Discussion} \label{discussion}

\subsection{Critical field $B_{c1}$ and complete flux expulsion} 

In Section \ref{section_Bc1_hmg}, we focused on a homogeneous narrow thin-film strip and reproduced well-established results, such as the free energy $G$ and the critical field $B_{c1}$. 
This not only served as a summary of basic knowledge but also familiarized us with our approach in dealing with inhomogeneous systems.
As commonly known, the critical field $B_{c1}$ for a homogeneous film is described by Eq.~(\ref{Bc1_hmg}), which is nearly independent of material parameters. 
Consequently, the only viable method to increase $B_{c1}$ was believed to be reducing the film's width $W$, as discussed in previous studies. 
However, this constraint is valid solely for homogeneous films, as we demonstrated in Section \ref{section_Bc1_inhmg}.

In Section \ref{section_Bc1_inhmg}, we developed a formulation to calculate the free energy $G=\epsilon - \mu_z B$ of an inhomogeneous film. 
Specifically, we derived expressions for the self-energy $\epsilon$ given by Eqs.~(\ref{self_energy_inhmg}) and (\ref{self_energy_inhmg_edges}). 
Additionally, the magnetic potential contribution $U=-\mu_z B$ was evaluated using Eq.~(\ref{MM_inhmg}).
Employing these formulations, we investigated the effects of a quadratic $\Lambda(x)$ distribution on $G$ and $B_{c1}$ as an illustrative example [see also Fig.~\ref{fig3} (a)]. 
Our findings, depicted in Figure \ref{fig5}, revealed that inhomogeneities, such as having a shorter $\Lambda$ in the middle of the strip, can significantly enhance $B_{c1}$.

One significant application of these findings is the achievement of complete flux expulsion in superconducting films. 
Previous studies have demonstrated that when the ambient magnetic flux is smaller than the critical field ($B_{c1}$), it is possible to fully expel the flux from the film. 
As the critical field increases, the film becomes more resistant to flux trapping. 
Traditionally, the only method to increase $B_{c1}$ and promote flux expulsion in homogeneous films was to narrow the strip. 
However, the introduction of engineered inhomogeneities provides an alternative approach to enhance $B_{c1}$ and make the film robust against flux trapping. 
This advancement offers greater flexibility in designing superconducting devices with improved flux expulsion capabilities. 
Furthermore, these findings may offer insights into the flux expulsion phenomenon observed in spatial temperature gradients, which is utilized to enhance the quality factor of particle accelerator cavities~\cite{2014_Romanenko, 2016_Huang, 2016_Posen, 2021_Miyazaki, 2021_Ooi}. 

.

It is important to note that even in the absence of vortices and at low microwave frequencies compared to the superconducting gap, a finite surface resistance can arise due to thermally excited quasiparticles~\cite{MB, AGK}. 
This issue is further complicated by various pair-breaking mechanisms originating from the current and resonator materials~\cite{2011_Fominov, 2012_Proslier, 2012_Clem_Kogan, 2014_Gurevich, 2016_Semenov, 2017_Gurevich_Kubo, 2019_Kubo_Gurevich, 2021_Herman, 2020_Kubo_1, 2022_Kubo, 2022_Gurevich_Kubo_Sauls} and the nonequilibrium dynamics of quasiparticles and phonons triggered by sub-gap and pair-breaking photon~\cite{2019_Kubo_Gurevich, 2022_Gurevich_Kubo_Sauls, 2012_Visser, 2014_Visser, 2015_Withington, Barends, Naruse}. 
Recent research has explored strategies to reduce dissipation by engineering the quasiparticle density of states (DOS), 
including methods such as applying dc~\cite{2014_Gurevich, 2020_Kubo_1} or rf~\cite{2014_Gurevich, 2019_Kubo_Gurevich} currents, introducing sparse magnetic impurities~\cite{2017_Gurevich_Kubo, 2019_Kubo_Gurevich}, modifying subgap states through surface processing~\cite{2017_Gurevich_SUST, 2020_Kubo_1, 2022_Kubo}, or optimizing proximity effects between superconductors and metallic suboxides or hydrides~\cite{2017_Gurevich_Kubo, 2019_Kubo_Gurevich}. 
To optimize the quality factor, it is crucial to not only achieve a vortex-free state but also address the engineering of the DOS, as both factors play complementary roles in achieving low dissipation and high-performance superconducting devices.

\subsection{Critical current enhancement and diode effect} 

In Sec.~\ref{section_Ic_formulation}, we developed a formulation to evaluate the critical current in an inhomogeneous narrow thin-film strip. 
In the presence of an inhomogeneous $\Lambda(x)$ distribution, 
the sheet bias-current density deviates from uniformity and can be described by Eq.(\ref{inhmg_J}). Consequently, the contribution of the bias current to the free energy, denoted as $\Delta G$, is given by Eq.~(\ref{delta_G_inhmg}).

In Section~\ref{section_critical_current_inhmg}, our focus was specifically on examining the influence of a quadratic $\Lambda(x)$ distribution on the critical current and determining whether the introduction of inhomogeneity could lead to an enhancement in the critical current. 
Our findings led us to the conclusion that, when the objective is to increase the critical current, it is more advantageous to employ an inhomogeneous film with a reduced $\Lambda$ value in the middle of the strip, as opposed to a homogeneous film. 
The enhancement in critical current is attributed to the current suppression effect at the edges induced by the inhomogeneous superfluid density distribution. This mechanism shares similarities with the superheating field enhancement studied in inhomogeneous structures employed in particle accelerator resonators~\cite{2006_Gurevich, 2014_Kubo, 2015_Gurevich, 2017_Liarte_SUST, 2017_Kubo_SUST, 2019_Sauls, 2021_Kubo}.
Future work will involve comparing our findings with microscopic calculations of the depairing current in an inhomogeneous strip. This analysis will provide a more comprehensive understanding of the critical current behavior in such systems.

In Sec.~\ref{section_diode}, we explored the impact of a left-right asymmetric $\Lambda(x)$ distribution, described by Eqs.~(\ref{linear_Lambda}) and (\ref{linear_F}), on the superconducting diode effect. 
By deriving the expression for the free energy, Eq.~(\ref{G_linear}), we were able to determine the critical current, as depicted in Fig.~\ref{fig9} (b). 
The critical current depends on the interplay between the left-right asymmetric edge barriers and the left-right asymmetric sheet bias-current distribution. 
Furthermore, we analyzed the behavior of the superconducting diode quality parameter $\eta$ with respect to the inhomogeneity parameter $\alpha$ and the external field $B$, as presented in Fig.~\ref{fig9} (c). 
Our results revealed that increasing the inhomogeneity parameter and the external field led to an enhancement of $\eta$, showcasing the potential of engineering inhomogeneous $\Lambda$ distributions to control and manipulate the superconducting diode effect.

Any left-right asymmetric $\Lambda(x)$ distribution has the potential to induce the superconducting diode effect. One approach to achieve this is by creating a temperature gradient along the $x$ direction. This method offers a novel way to implement devices that exhibit the superconducting diode effect.


\begin{acknowledgments}
I am sincerely grateful to all those who generously supported my extended paternity leave, spanning from April 2021 to December 2023, enabling me to prioritize and enjoy invaluable time with my family~\cite{2023_Kubo_ikuji}. 
This work was supported by CASIO Science Promotion Foundation 34th Research grant (No. 22) and Toray
Science Foundation (No. 19-6004). 
\end{acknowledgments}

\appendix

\section{Vortex self-energy in an inhomogeneous superconductor} \label{appendix1}

We extend the method developed by Kogan~\cite{1994_Kogan, 2007_Kogan} for a homogeneous superconductor to study the behavior of a vortex in an inhomogeneous superconductor. 
In this case, we consider a superconductor with a nonuniform $\lambda ({\bf r})$. 
The self energy of a vortex can be divided into two parts.
\begin{eqnarray}
&&\epsilon = \epsilon_{\rm out} + \epsilon_{\rm in} , \label{epsilon1} \\
&&\epsilon_{\rm out} = \int dV_{\rm out} \frac{B^2}{2\mu_0} , \\
&&\epsilon_{\rm in} = \int dV_{\rm in} \biggl( \frac{B^2}{2\mu_0} + \frac{\mu_0}{2} \lambda^2 j^2 \biggr) .
\end{eqnarray}
Here, $\epsilon_{\rm out}$ is the magnetic energy outside the material, and $\epsilon_{\rm in}$ is the sum of the magnetic energy inside the material and the kinetic energy of the condensate.

Let us start from $\epsilon_{\rm out}$. 
Since ${\rm rot} {\bf B}=0$ in the vacuum, we can introduce a scalar potential $\varphi$, 
which gives ${\bf B}=\nabla \varphi$. 
Also, ${\rm div} {\bf B}$ yields $\nabla^2 \varphi=0$. 
Then, we get 
\begin{eqnarray}
\epsilon_{\rm out} 
= \frac{1}{2\mu_0} \int d{\bf S}_{\rm out}\cdot (\varphi \nabla \varphi) .
\end{eqnarray}
Here, $B^2 = (\nabla\varphi)^2=\nabla\cdot (\varphi\nabla\varphi)$ is used. 
Next, we calculate $\epsilon_{\rm in}$ using the Maxwell-London equation in an inhomogeneous superconductor, ${\bf B} + \mu_0 {\rm rot}(\lambda^2 {\bf j}) =0$ or
\begin{eqnarray}
{\bf B} + \lambda^2 {\rm rot}\, {\rm rot} {\bf B} + (\nabla \lambda^2) \times {\rm rot} {\bf B} =0 , \label{MLsc}
\end{eqnarray}
which is valid outside the vortex core. 
We find the following relation: 
\begin{eqnarray}
&&\nabla\cdot [(\lambda^2{\bf B}) \times {\rm rot}{\bf B} ] \nonumber \\
&=& {\rm rot}{\bf B} \cdot {\rm rot} (\lambda^2 {\bf B}) - ({\rm rot}\,{\rm rot} {\bf B})\cdot (\lambda^2 {\bf B}) \nonumber \\
&=& {\rm rot}{\bf B} \cdot {\rm rot} (\lambda^2 {\bf B}) + B^2 + [(\nabla \lambda^2) \times {\rm rot}{\bf B}]\cdot {\bf B} \nonumber \\
&=& \lambda^2 ({\rm rot}{\bf B})^2 + B^2 
= \mu_0^2 \lambda^2 j^2 + B^2.
\end{eqnarray}
Here, Eq.~(\ref{MLsc}) is used from the second line to the third line, 
and ${\rm rot}{\bf B} = \mu_0 {\bf j}$ is used in the last line. 
Hence, $\epsilon_{\rm in}$ can be written as
\begin{eqnarray}
\epsilon_{\rm in}  
&=& \frac{1}{2\mu_0} \int d{\bf S}_{\rm in}\cdot [(\lambda^2{\bf B}) \times {\rm rot}{\bf B}] 
= \epsilon_{\rm in}^{\rm (C)} + \epsilon_{\rm in}^{\rm (V)} .
\end{eqnarray}
Here, $\epsilon_{\rm in}^{\rm (C)}$ and $\epsilon_{\rm in}^{\rm (V)}$ are the contributions from the core surface and the vacuum-superconductor interface, respectively:  
\begin{eqnarray}
\epsilon_{\rm in}^{\rm (C)}  
&=& \frac{1}{2} \int d{\bf S}_{\rm in}^{\rm (C)} \cdot [(\lambda^2{\bf B}) \times {\bf j}] , \\
\epsilon_{\rm in}^{\rm (V)} 
&=& \frac{1}{2} \int d{\bf S}_{\rm in}^{\rm (V)}\cdot [(\lambda^2 \nabla \varphi ) \times {\bf j}] \nonumber \\
&=& -\frac{1}{2} \int d{\bf S}_{\rm in}^{\rm (V)}\cdot [ (\nabla \lambda^2)\times {\bf j} + \lambda^2 {\rm rot} {\bf j} ] \varphi . 
\end{eqnarray}
In the last line, ${\rm rot} (\lambda^2 \varphi {\bf j}) = \nabla((\lambda^2 \varphi) \times {\bf j} + (\lambda^2 \varphi {\rm rot}{\bf j}$ and $\nabla((\lambda^2 \varphi)= \lambda^2 \nabla\varphi + \varphi(\nabla\lambda^2)$ are used. 
Using $d{\bf S}_{\rm in}^{\rm (V)}=-d{\bf S}_{\rm out}$, Eq.~(\ref{epsilon1}) results in 
\begin{eqnarray}
\epsilon &=& \epsilon_{\rm out} + \epsilon_{\rm in} \nonumber \\
&=&  \frac{1}{2\mu_0} \int d{\bf S}_{\rm out} \cdot [ {\bf B} + (\nabla \lambda^2)\times {\rm rot}{\bf B} \nonumber \\ 
&& + \lambda^2 {\rm rot}\, {\rm rot} {\bf B} ] \varphi + \epsilon_{\rm in}^{\rm (C)}  \nonumber \\
&=& \frac{\phi_0}{2\mu_0} [\varphi({\bf r}_{\rm ent}) -\varphi({\bf r}_{\rm ex}) ] + \epsilon_{\rm in}^{\rm (C)} 
\end{eqnarray}
$\varphi({\bf r}{\rm ent})$ and $\varphi({\bf r}{\rm ex})$ represent the potentials at the points where the vortex enters and exits, respectively.

Now, let's consider thin films positioned on the $xy$ plane. In the case of a thin film, we can neglect $\epsilon_{\rm in}^{\rm (C)}$, resulting in the following simplified expression
\begin{eqnarray}
\epsilon = -\frac{\phi_0}{\mu_0} \varphi({\bf r}_{\rm ex}).
\end{eqnarray}
Integrating $\mu_0 {\bf j} = {\rm rot}\,{\bf B}$ over the film thickness, 
we get 
\begin{eqnarray}
&&J_x 
= -\frac{2}{\mu_0}B_y(+0) = -\frac{2}{\mu_0} \partial_y \varphi(+0) , \\ 
&&J_y 
= \frac{2}{\mu_0}B_x(+0) = \frac{2}{\mu_0} \partial_x \varphi(+0) , 
\end{eqnarray}
On the other hand, we have the relations $J_x =\partial_y \Psi$ and $J_y = -\partial_x \Psi$. 
We get $\varphi(+0) = -(\mu_0/2) \Psi$. 
Then, finally we get
\begin{eqnarray}
\epsilon = \frac{\phi_0}{2} \Psi ({\bf r}_v).
\end{eqnarray}
which corresponds to the formula derived by Kogan for a homogeneous film.

\section{Derivations of $I_{\rm V, AV}^{(\pm)}$} \label{derivation}

The vortex-entry current $I_{\rm V}$ (or antivortex-entry current $I_{\rm AV}$) represents the current at which the edge barrier vanishes. 
This condition can be expressed as $G(X_p; I_{\rm V, AV})=0$, 
where $X_p$ denotes the peak position of the edge barrier, satisfying $G'(X_p; I_{\rm V, AV}) = 0$.

To get $I_{\rm V}^{(+)}$ and $I_{\rm AV}^{(-)}$, we expand $G(X)$ around the left edge $X/W=-1/2+\delta$ ($\delta \ll 1$). 
We find
\begin{eqnarray}
&&\frac{G(X)}{\epsilon_0} =\frac{1}{F(-W/2)} \biggl[ \ln\frac{2\delta}{\xi_L/W} \mp (\tilde{B}_L +\tilde{I}) \delta \biggr]
\end{eqnarray}
for a vortex (an antivortex) in the current $I\ge 0$ ($I \le 0$). 
To get $I_{\rm V}^{(-)}$ and $I_{\rm AV}^{(+)}$, we expand $G(X)$ around the right edge $X/W=1/2-\delta$.  
We find
\begin{eqnarray}
&&\frac{G(X)}{\epsilon_0} =\frac{1}{F(W/2)} \biggl[ \ln\frac{2\delta}{\xi_R/W} \pm (\tilde{B}_R +\tilde{I}) \delta \biggr]
\end{eqnarray}
for a vortex (an antivortex) in the current $I\le 0$ ($I \ge 0$). 
Here, $\tilde{I}$ and $\tilde{B}_{L, R}$ are given as follows.

For a homogeneous film ($F=1$ and $\xi_L=\xi_R=\xi$), 
we have
\begin{eqnarray}
&&\tilde{I} = \frac{I}{I_0} , \\
&&\tilde{B}_{L, R} = \pm \frac{4B}{B_{\phi}} .
\end{eqnarray}
For an inhomogeneous film with a quadratic distribution of $\Lambda$ (see Fig.~\ref{fig5}), 
we have 
\begin{eqnarray}
&&\tilde{I} = \frac{I}{I_0} \frac{4-\gamma}{4} \frac{\sqrt{\gamma/(4-\gamma)}}{\tan^{-1}\sqrt{\gamma/(4-\gamma)}} ,  \\
&&\tilde{B}_{L, R} = \pm \frac{4B}{B_{\phi}} .
\end{eqnarray}
For an inhomogeneous film with a left-right asymmetric distribution of $\Lambda$ (see Fig.~\ref{fig8}), 
we have 
\begin{eqnarray}
&&\tilde{I} = \frac{I}{I_0} \frac{2\alpha}{\ln [(1+\alpha)/(1-\alpha)]} , \\
&&\tilde{B}_{L, R} = \frac{2(B/B_{\phi})}{\alpha \tanh^{-1}\alpha} \biggl( 2\alpha - (1\mp \alpha) \ln\frac{1+\alpha}{1-\alpha} \biggr) .
\end{eqnarray}
Note that, taking $\gamma \to 0$ and $\alpha \to 0$, we can reproduce the results for the homogeneous case from the inhomogeneous cases. 

The condition of the disappearance of the edge barrier yields 
\begin{eqnarray}
\tilde{I}_{\rm V}^{(\pm)} = \frac{2W}{e \xi_{L, R}} \mp \tilde{B}_{L, R} , \\
\tilde{I}_{\rm AV}^{(\pm)} = \frac{2W}{e \xi_{R, L}} \mp \tilde{B}_{R, L} , 
\end{eqnarray}
which immediately yield Eqs.~(\ref{I_V_hmg}) and (\ref{I_AV_hmg}), 
Eqs.~(\ref{I_V_quadratic}) and (\ref{I_AV_quadratic}), and Eqs.~(\ref{I_V_linear}) and (\ref{I_AV_linear}).

\section{$B_{\rm stop}$ in inhomogeneous films} \label{appendix_Bstop}

As discussed in Section~\ref{section_Ic_hmg}, the linear dependence of $I_c$ on $B$ breaks down when $B \gtrsim B_{\rm stop}$. At this critical magnetic field value, a free energy minimum appears at a point $X_s$ where $G'(X_s; I_c, B_{\rm stop})=0$, and vortices come to a stop at this minimum~\cite{Clem_Mawatari, Ilin_Vodo}. These vortices, carrying a current, then impact the barrier and modify the behavior of $I_c$.

\begin{figure}[tb]
   \begin{center}
   \includegraphics[width=0.49\linewidth]{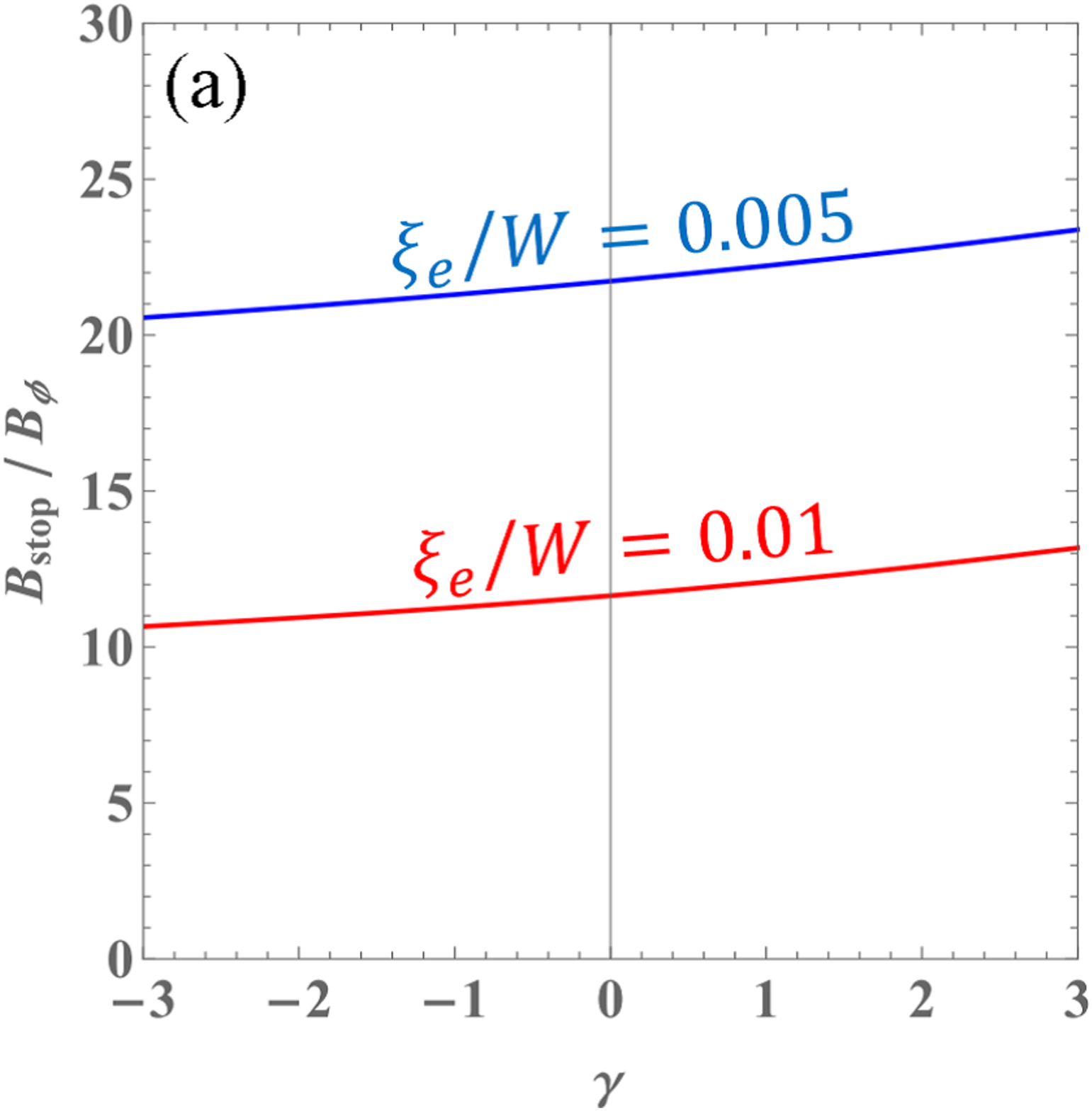}
   \includegraphics[width=0.49\linewidth]{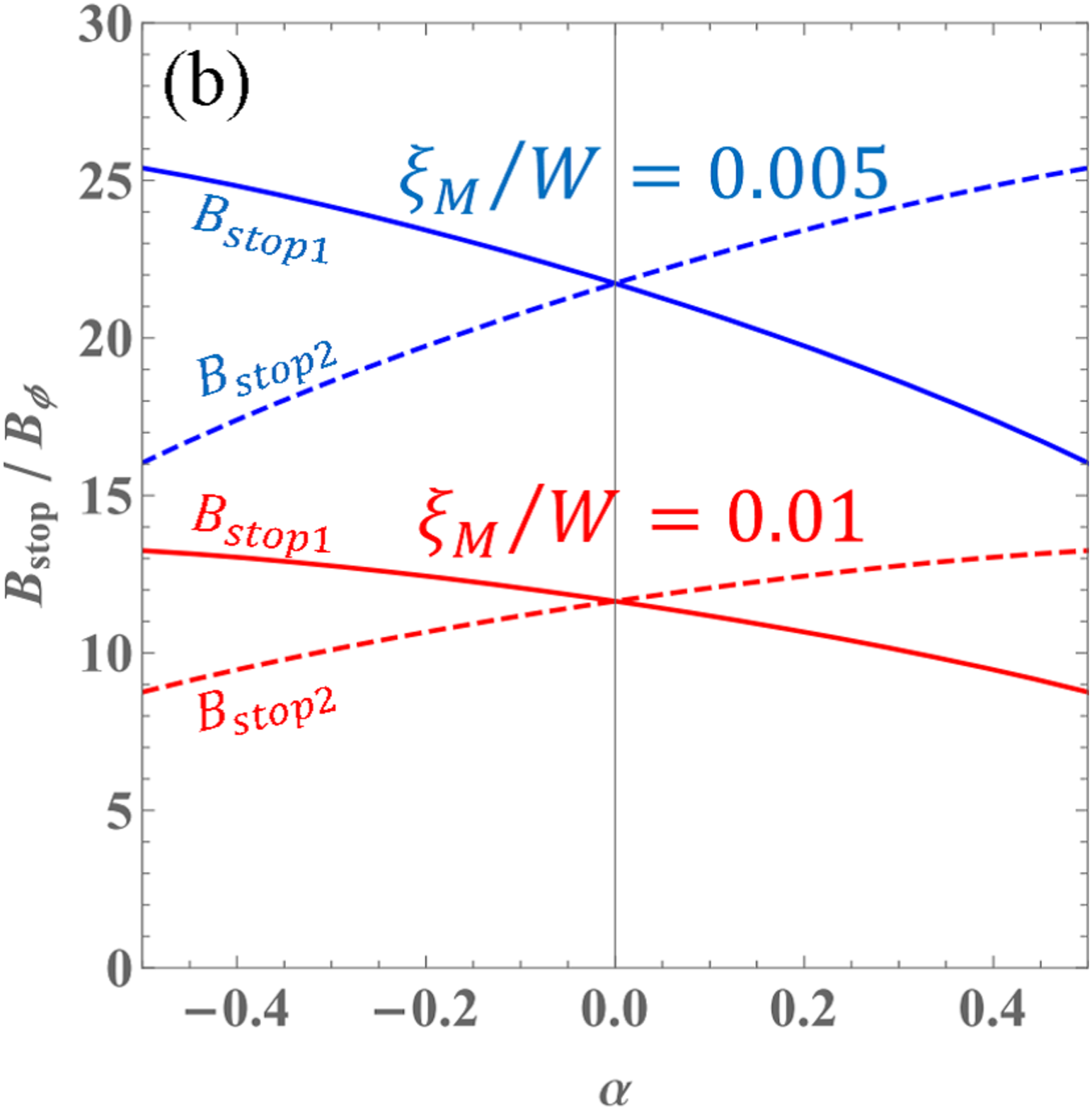}
   \end{center}\vspace{0 cm}
   \caption{
$B_{\rm stop}$ of inhomogeneous narrow thin-film strips as a function of the inhomogeneity parameter.
(a) Quadratic distribution of $\Lambda(x)$ (see Fig.~\ref{fig3}). The coherence length at the edges is denoted as $\xi_e=\xi|{x=\pm W/2}$.
(b) Left-right asymmetric distribution of $\Lambda(x)$ (see Fig.~\ref{fig8}). 
Here, we consider the inhomogeneity to arise from variations in impurity concentration. In this case, the coherence length at the edges can be calculated as $\xi_L=\xi_M/F(-W/2)=\xi_M/\sqrt{1-\alpha}$ and $\xi_R=\xi_M/F(W/2)=\xi_M/\sqrt{1+\alpha}$, where $\xi_M=\xi|_{x=0}$ is the coherence length at the middle of the strip. 
   }\label{fig10}
\end{figure}

The critical value $B_{\rm stop}$ is determined by the condition for the existence of a solution $X_s$ that satisfies $G'(X_s; I_c, B_{\rm stop})=0$. For a film with a quadratic distribution of $\Lambda(x)$ (as shown in Fig. 3), we analyze this condition using the expression of $G$ given by Eq.~(\ref{G_quadratic}). The dependence of $B_{\rm stop}$ on $\gamma$ is illustrated in Fig.~\ref{fig10} (a). Moreover, when the film is homogeneous ($\gamma=0$), the calculated value of $B_{\rm stop}$ coincides with the result obtained from Eq.~(\ref{Bstop_hmg}).

On the other hand, for a film with a left-right asymmetric distribution of $\Lambda(x)$ (as depicted in Fig. 8), we consider the condition using $G$ given by Eq.~(\ref{G_linear}), resulting in two distinct critical values: $B_{\rm stop1}$ and $B_{\rm stop2}$ for penetration from the left edge and right edge, respectively. The linear dependence on $B$ holds within the ranges $-B_{\rm stop2} < B < B_{\rm stop1}$ for $I_c^{(+)}$ and $-B_{\rm stop1} < B < B_{\rm stop2}$ for $I_c^{(-)}$.

\end{document}